\documentclass[12pt]{article}
\usepackage{amsfonts}
\usepackage{amsmath}
\usepackage{amssymb}
\usepackage{graphicx}

\begin{document}

\title{A Cumulative Framework for Solid Deformation}
\author{Lev Steinberg \\
Independent Researcher, Lexington, Kentucky, USA}
\date{}
\maketitle

\begin{abstract}
A cumulative variational framework for solid deformation is proposed. Its central organizing
principle is the recursive combination of three structural components of one
material body: a compatible Cosserat component, a distributed-incompatibility
component, and a defect-source component. At any prescribed external
condition these components constitute a \emph{combined material state}. They
are governed under common physical boundary and initial conditions and are
coupled through one cumulative energy and the associated field equations.
They are not separate boundary-value problems and need not appear in a
prescribed temporal sequence.

The compatible component is described by a coframe and rotational connection,
the defect component by distributed torsion and curvature, and the source
component by independently variable defect fields and their geometric
nonclosure characteristics. All fields are represented materially on the
common reference body. Their recursive geometric dependence is incorporated
directly into the variation of a single stored energy, yielding a coupled
Euler--Lagrange system without introducing separate phenomenological reaction
fields.

The cumulative character of the theory concerns the response of the combined
state to variation of externally prescribed conditions. Introducing a
loading-path parameter $\lambda$, the state is written
\begin{equation}
\mathcal Z(\lambda)=\bigl(\mathcal Z_c(\lambda),\mathcal Z_d(\lambda),
\mathcal Z_s(\lambda)\bigr).
\end{equation}
Its components may coexist in the initial or current state, while their
relative participation may change through path-dependent activation and
retention as $\lambda$ varies. For the special case of an initially perfect
body, loss of stability may produce successive branches with defect and then
source participation. Such a sequence is a bifurcation property of the
selected initial state and constitutive energy, rather than a defining
hierarchy of the framework.

A one-dimensional specialization illustrates structural bifurcation,
residual deformation and residual stress produced by retained structure,
dependence on the initial combined state, Bauschinger-type directional
memory, parametric cumulative hysteresis, and changes of observable tangent
stiffness associated with structural stabilization. A spatial extension
introduces defect and source gradient terms and yields a size-dependent
structural stability operator. For the stable quartic energy considered, the
gradient terms regularize nonuniform perturbations but do not by themselves
generate a finite-wavelength localization instability on an already stable
three-component branch. The examples thereby distinguish consequences of the
cumulative structure and energy from phenomena requiring additional
constitutive evolution laws.
\end{abstract}

\tableofcontents

\noindent\textbf{Keywords:} cumulative framework; solid deformation; Cosserat
continuum; mesoelasticity; distributed defects; differential forms;
recursive variational structure; configurational mechanics.

\newpage

\section{Introduction}

Classical continuum theories provide the natural starting point for the
mechanics of solids. Classical elasticity describes compatible deformation,
while the Cosserat brothers enlarged continuum kinematics by assigning
independent rotational degrees of freedom to material points \cite%
{Cosserat1909}. Modern microcontinuum formulations provide broader classes
of internal kinematic fields \cite{Eringen1999}. These developments
demonstrate that the macroscopic deformation of a body need not exhaust its
material structure.

The geometric description of incompatibility developed along a complementary
line. Kondo introduced non-Riemannian geometry into the continuum
description of yielding and defects \cite{Kondo1952}; Bilby, Bullough, and
Smith described continuous dislocation distributions geometrically \cite%
{Bilby1955}; and Kr\"{o}ner subsequently developed a systematic continuum
theory of defects and incompatibility \cite{Kroner1981}. In the
metric-compatible Cosserat, or Riemann--Cartan, setting adopted here,
translational incompatibility associated with distributed dislocations is
represented naturally by torsion, whereas rotational incompatibility
associated with distributed disclinations is represented by curvature.
Nonmetricity is not introduced. The corresponding Bianchi identities express
geometric nonclosure relations for these defect fields. The
differential-form and manifold language used below follows standard
treatments including Flanders \cite{Flanders1963}, Edelen \cite{Edelen1985},
and Choquet--Bruhat, DeWitt-Morette, and Dillard-Bleick \cite%
{ChoquetBruhat1982}.

Gauge and differential-geometric formulations provide further precedents.
Edelen treated prescribed material dislocation-density two-forms as material
Cartan torsion within a gauge theory of solids \cite{Edelen1989}, while Duan
and Huang formulated a four-dimensional nonlinear defect continuum in which
dislocation and disclination densities and currents are represented through
torsion and curvature and corresponding continuity relations follow from the
Bianchi identities \cite{DuanHuang1990}. Recent moving-frame formulations
organize generalized continua geometrically through moving frames and
reductions of the relevant structural group \cite{EckerKolev2026}.
Configurational mechanics provides a complementary description of material
driving forces, from Eshelby's elastic singularity force \cite{Eshelby1951}
to systematic continuum formulations \cite{Gurtin2000,Maugin1993}. Recent
work also treats regular bulk and interfacial defects within a unified
configurational framework \cite{DiStefano2026}.

Generalized-continuum, gradient, and internal-variable theories provide a
second important background. Gradient theories introduce additional spatial
structure and characteristic length scales relevant to localization and size
effects \cite{Aifantis1992,FleckEtAl1994,Forest2009,Gurtin2004}, while
classical theories of cyclic plasticity and hardening enlarge the material
state through evolving internal variables \cite%
{ArmstrongFrederick1966,Chaboche1986,LemaitreChaboche1990}. Rubin and
Panteghini provide a recent example in which macro- and micro-continua
retain distinct but interacting responses; their micro-continuum is
represented by a deformable director triad with additional director-momentum
and director-gradient effects \cite{RubinPanteghini2026}.

Physical observations likewise indicate that internal material structure may
evolve and retain information not contained in the instantaneous macroscopic
strain. Guyer and Johnson emphasized strong elastic nonlinearity,
hysteresis, and memory associated with mesoscopic internal contacts and
linkages in heterogeneous solids \cite{GuyerJohnson1999}. Kuhlmann-Wilsdorf
related work hardening to the generation, rearrangement, and organization of
dislocation structures into accessible low-energy configurations \cite%
{KuhlmannWilsdorf1989,KuhlmannWilsdorf1998,KuhlmannWilsdorf1999}. These
mechanisms are not identified here with the geometric fields introduced
below. Rather, they illustrate a more general fact important for the present
theory: deformation may alter the internal material structure, and structure
accumulated during an earlier part of the loading history may influence the
subsequent response.

Despite these developments, a basic structural question remains. Can
compatible deformation, distributed incompatibility, and defect-source
structure be represented as distinct but recursively related components of a
single material body, governed by common physical boundary and initial
conditions, while retaining the geometric and variational identity of each
component? A further question arises when several structural responses are
simultaneously available: what material-response principle determines their
preferential participation as externally affecting conditions change? The
present paper addresses these questions by constructing a cumulative
variational theory in which the three structural components may coexist,
their activation is path-dependent, and the response is governed by a
material response strategy. The compatible component is characterized by the
Cosserat coframe and rotational connection, distributed incompatibility by
torsion and curvature, and the defect-source component by higher-order
nonclosure characteristics termed hyper-torsion and hyper-curvature. Their
precise geometric definitions and recursive relations are developed below.
The theory concerns deformation prior to macroscopic fracture;
fracture-surface formation is regarded as a subsequent structural transition
beyond the present scope.

\medskip \noindent\textbf{Notation and Conventions.}

Spatial Euclidean indices are $a,b,c=1,2,3$, reference material indices are $%
A,B,C=1,2,3$, and internal Cosserat indices are $i,j,k=1,2,3$. Structural
subscripts are labels rather than hierarchical levels: $c$ denotes the
compatible structural component, $d$ the distributed-incompatibility
structural component, and $s$ the defect-source structural component. The
terms \emph{hyper-torsion} and \emph{hyper-curvature} denote the geometric
nonclosure characteristics associated with the defect-source component;
their precise definitions are introduced below.

Throughout this work, $\Omega^{p}(M)$ denotes the space of smooth
differential $p$-forms on the material manifold $M$, while $\Omega^{p}(M,%
\mathfrak{so}(3))$ denotes the space of $\mathfrak{so}(3)$-valued
differential $p$-forms. Exterior differentiation is denoted by $d$ and the
wedge product by $\wedge$. The covariant exterior derivative associated with
a specified rotational connection is denoted by $D$, with a structural
subscript when the relevant connection must be distinguished. The connection 
$\omega^i{}_j$ is skew-symmetric, $\omega_{ij}=-\omega_{ji}$.
Coordinate-free, the oriented Cosserat frames form a principal $SO(3)$%
-bundle over the material manifold; bundle language is used only when needed
to distinguish transformation laws and carrier spaces. All calculations are
subsequently written in a local orthonormal frame.

\medskip \noindent\textbf{Proposed Cumulative Framework.}

The present paper addresses this problem by proposing a cumulative framework
in which compatible deformation, distributed incompatibility, and
defect-source structure are treated as recursively related components of a
single material structure. The distinction between these components is
retained because they describe geometrically different characteristics of the
material body, but they are not regarded as independent theories or as
separate boundary-value problems.

At a prescribed value of an externally controlled parameter $\lambda$, the
material body is characterized by the \emph{combined state}
\begin{equation}
\mathcal Z(\lambda)=\bigl(\mathcal Z_c(\lambda),\mathcal Z_d(\lambda),
\mathcal Z_s(\lambda)\bigr),
\end{equation}
where $\mathcal Z_c$, $\mathcal Z_d$, and $\mathcal Z_s$ denote respectively
the compatible, distributed-incompatibility, and defect-source components.
Any of these components may already be present in the initial state, and
their simultaneous presence does not imply a temporal hierarchy.

The term \emph{cumulative} refers instead to the response of this combined
state as the externally affecting conditions vary along a path parametrized
by $\lambda$. The parameter $\lambda$ is a loading-path parameter and need
not represent physical time. The resulting family $\{\mathcal Z(\lambda)\}$
describes the cumulative response of the material body.

A \emph{material response strategy} specifies how the relative participation
of the structural components may change along this path. For an initially
perfect body, the compatible component may respond first; continued change of
the external conditions may activate distributed incompatibility, and
further change may activate defect-source structure. Symbolically, one
possible response is
\begin{equation}
(\mathcal Z_c,0,0)\longrightarrow(\mathcal Z_c,\mathcal Z_d,0)
\longrightarrow(\mathcal Z_c,\mathcal Z_d,\mathcal Z_s).
\end{equation}
This ordering expresses a possible \emph{path-dependent activation} of
structural components. It is not a universal temporal sequence and does not
imply that the three components constitute separate levels of material
existence. For a nonvirgin initial state, two or all three components may
participate from the beginning.

The recursion is geometric. The distributed-incompatibility component is
represented relative to the retained compatible structure, while the
defect-source component is represented relative to the retained
compatible--defect structure. Consequently, activation of an additional
component does not replace the preceding structure; it enlarges the combined
state. Upon reversal of the external parameter, previously activated
components need not disappear or retrace their previous states. Their
retention provides a geometric mechanism for residual deformation, residual
stress, directional memory, and path-dependent macroscopic response.

All structural components are subject to the same physical boundary and
initial conditions of the material body. Their recursive combination under
these common conditions, together with the cumulative energy and its
variational equations, constitutes the proposed cumulative framework for solid
deformation.

\section{Fundamental Principles of Cumulative Mechanics}

This section collects the defining principles that organize the proposed cumulative
mechanics before the detailed variational field equations are introduced. The
combined structure specifies which geometric components are present at a given
state; the material response strategy describes their path-dependent participation;
stability provides a qualitative criterion for loss of local accessibility of a
followed equilibrium branch; and the recursive relations specify how the components
are geometrically coupled within the same Cosserat body.

\subsection{Combined Structure of the Material Body}

The framework considers a Cosserat material body $\mathcal{B}$ located in the
ambient physical Euclidean space $\mathcal{E}^{3}$. Each material point
possesses the translational and orientational characteristics of a Cosserat
continuum. The internal material microstructure of the body is described
geometrically through three related structural components: a compatible
Cosserat component, a distributed-incompatibility component, and a
defect-source component.

These components provide progressively enriched descriptions of the internal
microstructure of the same material body. They do not represent different
bodies and, by themselves, do not imply a temporal sequence of material
states. At a given state they may coexist and together constitute the \emph{%
combined internal structure of the material body}. The cumulative character
of the theory arises subsequently through their path-dependent
participation, activation, and retention according to the material response
strategy.

\begin{figure}[htbp]
\centering
\includegraphics[width=0.90\textwidth]{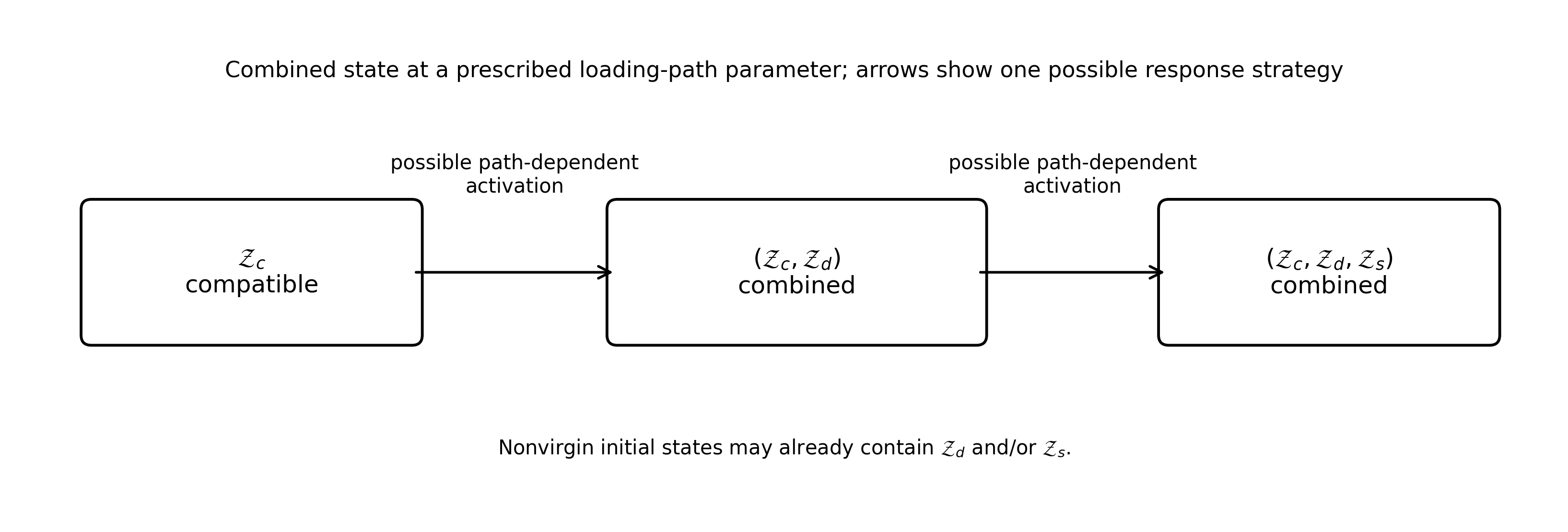}
\caption{Combined structural components and one possible path-dependent activation
strategy. At each value of the loading-path parameter the boxes represent a combined
state; the arrows indicate possible activation and not a universal temporal sequence.}
\label{fig:fundamental-components}
\end{figure}

Figure~\ref{fig:fundamental-components} summarizes one possible path-dependent
activation of the structural components. The arrows indicate a material-response
strategy for an initially compatible body and do not represent a compulsory
hierarchy; a nonvirgin initial state may already contain distributed-incompatibility
and/or defect-source structure.

The compatible component is described by 
\begin{equation}
\mathcal{Z}_{c}=(e_{c},\omega_{c};0,0),
\end{equation}
where $e_{c}$ is the Cosserat coframe and $\omega_{c}$ is the corresponding
rotational connection. If $D_{c}$ denotes the covariant exterior derivative
associated with $\omega_{c}$, compatibility is characterized by 
\begin{equation}
T_{c}:=D_{c}e_{c}=0,\qquad
\Omega_{c}:=d\omega_{c}+\omega_{c}\wedge\omega_{c}=0.
\end{equation}

The structure 
\begin{equation}
(e_{c},\omega_{c};T_{c}=0,\Omega_{c}=0)
\end{equation}
is the geometric representation of a compatible classical Cosserat material
space. Indeed, as shown in \cite{Steinberg2026Palatini}, a compatible
Cosserat motion is described by a placement $y$ and an independent
micro-orientation $R$, for which 
\begin{equation}
e_{c}=R^{-1}dy,\qquad \omega_{c}=R^{-1}dR.
\end{equation}
Consequently, 
\begin{equation}
T_{c}=D_{c}e_{c}=0,\qquad
\Omega_{c}=d\omega_{c}+\omega_{c}\wedge\omega_{c}=0.
\end{equation}
Conversely, locally on a simply connected compatible region, a flat
rotational connection and torsion-free coframe admit such a representation.
Thus vanishing torsion and curvature express compatibility without
eliminating the independent Cosserat micro-orientation. The quantities $%
T_{c} $ and $\Omega_{c}$ are therefore compatibility characteristics and not
additional independent state variables.

The distributed-incompatibility component describes the same Cosserat
material body when its internal microstructure is incompatible. It is
represented by 
\begin{equation}
\mathcal{Z}_{d}=(e_{d},\omega_{d};T_{d},\Omega_{d}),
\end{equation}
where $e_{d}$ and $\omega_{d}$ describe the local incompatible Cosserat
microstructure and 
\begin{equation}
T_{d}:=D_{d}e_{d} =de_{d}+\omega_{d}\wedge e_{d},\qquad
\Omega_{d}:=d\omega_{d}+\omega_{d}\wedge\omega_{d}
\end{equation}
are its distributed torsion and distributed curvature.

Thus $e_{d}$ and $\omega_{d}$ describe the local Cosserat microstructure,
whereas $T_{d}$ and $\Omega_{d}$ characterize, respectively, its distributed
translational and rotational incompatibility. In the absence of defect
sources they satisfy the Bianchi identities 
\begin{equation}
D_{d}T_{d}-\Omega_{d}\wedge e_{d}=0,\qquad D_{d}\Omega_{d}=0.
\end{equation}
Accordingly, $T_{d}$ and $\Omega_{d}$ are derived geometric characteristics
of the incompatible coframe and connection rather than additional
independent fields. Their surface integrals determine the corresponding
Burgers and Frank contents and provide the geometric representation of
continuously distributed dislocations and disclinations.

The passage from 
\begin{equation}
(e_{c},\omega_{c};0,0)
\end{equation}
to 
\begin{equation}
(e_{d},\omega_{d};T_{d},\Omega_{d})
\end{equation}
does not introduce a different material body. It represents an enrichment of
the internal geometric microstructure of the same Cosserat body from
compatible structure to distributed incompatible structure. This component
corresponds to the distributed-defect formulation developed in \cite%
{Steinberg2026Meso}.

The defect-source component describes the next order of internal
microstructure of the same Cosserat material body. It is represented by 
\begin{equation}
\mathcal{Z}_{s}=(T_{s},\Omega_{s};S,R),
\end{equation}
where $T_{s}$ is the source-associated torsion and $\Omega_{s}$ is the
source-associated curvature. These fields are independently variable. Relative
to the distributed-incompatibility structure, their higher-order nonclosure is
characterized by 
\begin{equation}
S:=D_{d}T_{s}-\Omega_{s}\wedge e_{d},\qquad R:=D_{d}\Omega_{s}.
\end{equation}

The three-form $S$ is termed \emph{hyper-torsion}, while $R$ is termed \emph{%
hyper-curvature}. They characterize, respectively, the translational and
rotational nonclosure of the independently variable defect fields $T_{s}$
and $\Omega_{s}$, and thereby describe the defect-source microstructure of
the material body. The term \emph{source} refers here to this higher-order
geometric nonclosure and does not by itself denote a temporal rate of defect
production.

When 
\begin{equation}
S=0,\qquad R=0,
\end{equation}
the source-level defect fields satisfy the corresponding closure relations.
This condition does not require 
\begin{equation}
T_{d}=0,\qquad \Omega_{d}=0.
\end{equation}
Consequently, a material body may possess distributed incompatibility
without possessing a nonvanishing defect-source structure.

The three structural components may therefore be summarized as 
\begin{equation}
(e_{c},\omega_{c};0,0),\qquad (e_{d},\omega_{d};T_{d},\Omega_{d}),\qquad
(T_{s},\Omega_{s};S,R).
\end{equation}
They represent, respectively, the compatible Cosserat microstructure, the
distributed-incompatibility microstructure, and the defect-source
microstructure of the same material body.

Their simultaneous presence defines the \emph{combined internal structure}
at a given material state. Combination alone does not imply a loading
history, an order of activation, or a succession of states. The cumulative
character of material behavior arises when these structural components
participate in the response through path-dependent activation and retention
according to the material response strategy. Thus the combined internal
structure specifies what structural components are present at a given state,
whereas the cumulative framework describes how material response develops
through those components along a loading path.

\subsection{Material Response Strategy and Cumulative Response}

The combined state does not by itself determine how the material
responds to a change of externally affecting conditions. Such a response
requires a material strategy specifying the priority with which the
available structural components participate in accommodating the imposed
change. For the class of materials considered in the present paper, we
postulate the strategic priority 
\begin{equation}
Z_1\prec Z_2\prec Z_3,
\end{equation}
where $Z_1$ denotes compatible deformation, $Z_2$
distributed-incompatibility evolution, and $Z_3$ defect-source evolution.
This ordering is a constitutive property of the material class considered
here and is not a universal requirement of the cumulative framework. Other
materials may possess a different priority or may admit coupled
participation of several structural components.

For simplicity, the quasistatic response is parametrized by the single
monotonically advancing scalar loading-path parameter $\lambda$ introduced
above. Let $P=P(\lambda)$ denote a scalar parameter characterizing the
prescribed external loading along this path. Depending on the boundary-value
problem, $P$ may represent an applied force or couple, a prescribed
displacement or rotation, or the amplitude of a prescribed combination of
external conditions. It is not a material-state variable. We distinguish 
\begin{equation}
\frac{dP}{d\lambda}>0
\end{equation}
as loading and 
\begin{equation}
\frac{dP}{d\lambda}<0
\end{equation}
as unloading. Thus $\lambda$ always advances along the response history;
loading and unloading are distinguished by the direction in which the
prescribed external loading changes, not by reversing the loading-path
parameter. A vanishing derivative may correspond to a turning point, a hold,
or a segment on which $P$ remains unchanged. For a general multidimensional
loading path, $P(\lambda)$ is understood as the scalar loading parameter
chosen to characterize the prescribed path.

The strategic priority applies to changes of the external conditions in
either direction. The material first accommodates an admissible incremental
change through $Z_1$. When the compatible response alone can no longer
provide the admissible continuation, $Z_2$ may participate; when the
combined $Z_1+Z_2$ response is insufficient or loses admissibility, $Z_3$
may participate. Activation of a higher component does not require the
preceding components to cease changing. The ordering $Z_1\prec Z_2\prec Z_3$
therefore specifies priority of participation rather than a temporal
succession of mutually exclusive states.

The same strategy governs unloading and reversal. A structural component
developed during loading may remain retained while its loading-path
increment vanishes over an initial portion of unloading. If further
unloading requires structural evolution, the corresponding $Z_2$ or $Z_3$
increment may become nonzero and may be directed oppositely to that
developed during loading. The strategy determines which structural mechanism
is given priority in accommodating the imposed change; it does not prescribe
the sign of the resulting structural increment.

Let $\mathcal{E}=W(\mathcal{Z})$ denote the stored internal energy of the
combined state. Along the actual quasistatic response path, 
\begin{equation}
d_{\lambda}\mathcal{E}>0
\end{equation}
represents internal-energy accumulation, whereas 
\begin{equation}
d_{\lambda}\mathcal{E}<0
\end{equation}
represents internal-energy release. These energetic signs are distinct from
the loading--unloading distinction defined by $dP/d\lambda$. Unloading
commonly permits energy release, but the two notions are not mathematically
identical. Nor is $d_{\lambda}\mathcal{E}$ identified with external work
when dissipation or other energetic channels are present.

Finally, the material strategy governs the actual incremental continuation
of the combined state and does not alter the geometric identities of the
structural variables or the Euler--Lagrange equations obtained from their
admissible virtual variations. Accordingly, $\delta$ and $d_{\lambda}$ have
different meanings: $\delta$ denotes a virtual variation at fixed $\lambda$,
whereas $d_{\lambda}$ describes passage between neighboring equilibrated
states along the actual loading path.

\subsection{Qualitative Stability and Structural Activation}

The material strategy introduced above is constitutive and is not derived
from stability. Stability has a different role: it identifies states at
which the presently followed equilibrium branch may cease to be locally
accessible, so that participation of additional structure becomes possible.
The analysis below is therefore qualitative and supportive of the cumulative
response construction rather than a derivation of its strategic priority.

Let $\Pi_{\lambda}$ denote the total potential corresponding to the
prescribed loading $P(\lambda)$ and evaluated about the cumulative
equilibrium state $\mathcal{Z}(\lambda)$. Stability with respect to
admissible distributed-incompatibility perturbations $\eta_d$ may be
characterized qualitatively by 
\begin{equation}
\mu_d(\lambda)=\inf_{\eta_d\neq0} \frac{\delta^2\Pi_{\lambda}[\eta_d,\eta_d]%
}{\|\eta_d\|^2}.
\end{equation}
A limiting condition 
\begin{equation}
\mu_d(\lambda_d)=0
\end{equation}
marks loss of local stability with respect to such perturbations. It does
not prescribe the material strategy or determine the subsequent branch; it
only identifies a state at which continuation of the presently followed
response may lose local stability.

On a retained compatible--distributed-incompatibility background, stability
with respect to admissible defect-source perturbations $\eta_s$ may likewise
be characterized by 
\begin{equation}
\mu_s(\lambda)=\inf_{\eta_s\neq0} \frac{\delta^2\Pi_{\lambda}[\eta_s,\eta_s]%
}{\|\eta_s\|^2},
\end{equation}
with a possible limiting condition 
\begin{equation}
\mu_s(\lambda_s)=0.
\end{equation}
For monotone loading from a perfect initial state, these limits may
accompany a realized response in which distributed incompatibility and
subsequently defect-source structure participate. The realized sequence,
however, is determined jointly by the initial state, loading path,
constitutive energy, boundary conditions, stability limits, and material
strategy; it is not itself postulated.

The same criterion applies along unloading and reverse-loading paths, but it
is then evaluated about the retained combined state produced by the prior
history. The corresponding limiting states therefore need not coincide with
those encountered during virgin loading. An initial unloading segment may,
for example, satisfy 
\begin{equation}
d_{\lambda}Z_1\neq0,\qquad d_{\lambda}Z_2=0,\qquad d_{\lambda}Z_3=0,
\end{equation}
with higher-level participation resuming only when continuation through the
retained response ceases to be admissible or stable. The subsequent
increments of $Z_2$ and $Z_3$ may oppose those developed during loading,
providing a natural setting for residual response, hysteresis, and
reverse-loading effects.

Energetically, the stored energy at a limiting state characterizes, for the
specified boundary-value problem, the capacity of the preceding response
branch. Such capacities depend on the constitutive energy, physical boundary
data, retained combined state, and limiting state; they are not additional
universal material constants. For an initially imperfect body, $\mathcal{Z}_d
$ and $\mathcal{Z}_s$ may already be nonzero, so the same stability
assessment is made about the retained combined state rather than about a
virgin perfect configuration.

\medskip Reduced theories are obtained as restrictions of the same combined
state and total energy. Setting 
\begin{equation}
T_s=\Omega_s=S=R=0
\end{equation}
reduces the formulation to the compatible--distributed-incompatibility
theory, provided the constitutive energy and its relevant derivatives reduce
consistently. Restricting the distributed-incompatibility component in
addition to its compatible baseline recovers the compatible Cosserat theory.
These are structural restrictions of the general formulation and not
temporal deactivation of components previously developed along a material
history.

Thus the material strategy and the stability criterion have distinct roles.
The strategy specifies the priority with which the structural components
participate in response to changing external conditions, whereas stability
determines whether the presently followed equilibrium branch remains locally
accessible. Neither condition alone determines the realized loading path.
The following reduced examples illustrate how their combination with the
constitutive energy and retained structural state can produce successive
activation, unloading, residual response, and reverse-loading behavior.

\subsection{Recursive Geometric Relations of the Combined Structure}

The combined internal structure is governed by recursive geometric relations
linking the compatible, distributed-incompatibility, and defect-source
components. For the compatible component, 
\begin{equation}
T_c=0,\qquad \Omega_c=0.
\end{equation}
For the distributed-incompatibility component, 
\begin{equation}
T_d^i=D_de_d^i,\qquad \Omega_d^i{}_{j} = d\omega_d^i{}_{j} +
\omega_d^i{}_{k}\wedge\omega_d^k{}_{j},
\end{equation}
and the corresponding Bianchi relations are 
\begin{equation}
D_dT_d^i-\Omega_d^i{}_{j}\wedge e_d^j=0, \qquad D_d\Omega_d^i{}_{j}=0.
\end{equation}
Here \emph{closure} means satisfaction of the corresponding Bianchi-type
geometric relation with zero source; it does not imply vanishing torsion or
curvature. Thus distributed incompatibility may be present even when these
closure relations are satisfied.

For the defect-source component, the independently variable defect fields $%
T_s$ and $\Omega_s$ are referred to the retained distributed-incompatibility
geometry, and their nonclosure is characterized by 
\begin{equation}
S^i = D_dT_s^i-\Omega_s^i{}_{j}\wedge e_d^j, \qquad R^i{}_{j} =
D_d\Omega_s^i{}_{j}.
\end{equation}
Accordingly, $S$ and $R$ measure the higher-order failure of closure of the
source-level defect fields and are termed hyper-torsion and hyper-curvature,
respectively. These relations express recursive geometric dependence among
the components of one combined material structure; they do not prescribe a
temporal sequence of activation.

\subsection{Cosserat Body Deformation and Structural Evolution}

The preceding section defined the combined internal structure of the
Cosserat material body. We now distinguish two related processes:
deformation of the body and evolution of its internal material structure.
Body deformation changes the placement of the material points and their
micro-orientation, whereas structural evolution changes the
distributed-incompatibility and defect-source components of the internal
structure. These processes are distinct but coupled through the material
response.

The compatible component 
\begin{equation}
\mathcal{Z}_{c}=(e_{c},\omega_{c};0,0)
\end{equation}
responds directly to the prescribed external physical loading. Its
deformation changes the placement of a material point $X\in B_0$ according
to 
\begin{equation}
x=\phi_c(X,\lambda),
\end{equation}
while preserving 
\begin{equation}
T_{c}=0,\qquad \Omega_{c}=0.
\end{equation}
Thus compatible deformation may continue while structural evolution is
active without losing its compatible geometric character.

In the quasistatic formulation considered here, for simplicity a single
scalar loading-path parameter $\lambda$ is used to parametrize the family of
neighboring equilibrated combined states. This does not restrict the
externally affecting conditions to a single physical control parameter; in
general they may be multidimensional, while $\lambda$ parametrizes a
prescribed path through their parameter space. It is not identified with
physical time or with a particular load magnitude. The same parameter
therefore indexes both the continuing compatible placement 
\begin{equation}
x=\phi_c(X,\lambda)
\end{equation}
and the structural state carried by the material point occupying that
position. The distributed-incompatibility and defect-source components may
accordingly be written spatially as 
\begin{equation}
\mathcal{Z}_{d}=\mathcal{Z}_{d}(x(X,\lambda),\lambda),\qquad \mathcal{Z}_{s}=%
\mathcal{Z}_{s}(x(X,\lambda),\lambda).
\end{equation}
Their material representation on $B_0$ is used in the field equations below.
The occurrence of $\lambda$ in the compatible placement does not make the
compatible fields structural-evolution variables; it only records that
compatible deformation may continue along the same quasistatic loading path.

For an infinitesimal loading-path increment $d\lambda$, we denote the
corresponding structural state differential by 
\begin{equation}
d_{\lambda}\mathcal{Z}_{\alpha} =D_{\lambda}\mathcal{Z}_{\alpha}\,d\lambda,%
\qquad \alpha=d,s.
\end{equation}
During an active structural-evolution stage, the accompanying increment of
compatible deformation is assumed small relative to the increment of the
active structural variables. The current compatible configuration may
therefore be treated, over that increment, as the instantaneous geometric
background for the corresponding quasistatic structural problem. This scale
separation belongs to the quasistatic specialization of the cumulative
theory and is distinct from the material-response strategy introduced below.

For the distributed-incompatibility component, 
\begin{equation}
\mathcal{Z}_{d} = (e_{d},\omega_{d};T_{d},\Omega_{d}),
\end{equation}
the coframe and rotational connection may change along the loading path, and
consequently their derived torsion and curvature also change. Their path
differentials satisfy 
\begin{equation}
d_{\lambda}T_{d}^{i} = D_{d}(d_{\lambda}e_{d}^{i}) +
(d_{\lambda}\omega_{d}^{i}{}_{j})\wedge e_{d}^{j},
\end{equation}
and 
\begin{equation}
d_{\lambda}\Omega_{d}^{i}{}_{j} = D_{d}(d_{\lambda}\omega_{d}^{i}{}_{j}).
\end{equation}
These relations describe evolution of the distributed incompatibility
already carried by the material body; they do not introduce an additional
deformation map.

The defect-source component 
\begin{equation}
\mathcal{Z}_{s} = (T_{s},\Omega_{s};S,R)
\end{equation}
evolves relative to the distributed-incompatibility geometry through 
\begin{equation}
S = D_{d}T_{s} - \Omega_{s}\wedge e_{d}, \qquad R = D_{d}\Omega_{s}.
\end{equation}
Hence source-structure evolution depends both on the independently variable
fields $T_{s}$ and $\Omega_{s}$ and on the evolving
distributed-incompatibility geometry $(e_{d},\omega_{d})$. This expresses
recursive geometric dependence without prescribing a temporal sequence of
activation.

\section{Variational Formulation of the Cumulative System}

The preceding section states the fundamental principles of cumulative mechanics:
the combined structure, the material response strategy, the qualitative role of
stability, the recursive geometric relations, and the distinction between body
deformation and structural evolution. We now formulate the coupled field problem.
The stress-field systems and constitutive relations are introduced first, and the
governing Euler--Lagrange equations then follow from stationarity with respect to
the admissible variations of the corresponding geometric fields. All structural
components remain subject to the common physical initial and boundary conditions
of the same material body.

\subsection{Stress-Field Systems}

In addition to the geometric fields introduced above, we postulate the
existence of stress-field systems for the compatible,
distributed-incompatibility, and defect-source structural components of the
material body. At this stage only their differential-form degrees are
specified; no constitutive relation or variational pairing is yet imposed.

For the compatible structural component we introduce the force-stress and
couple-stress two-forms 
\begin{equation}
\Sigma_{ci},\ M_c{}^i{}_j\in\Omega^2(B_0).
\end{equation}
For the distributed-incompatibility structural component we introduce 
\begin{equation}
\Sigma_{di},\ M_d{}^i{}_j\in\Omega^2(B_0), \qquad H_{di},\
O_d{}^i{}_j\in\Omega^1(B_0).
\end{equation}
For the defect-source structural component we introduce 
\begin{equation}
H_{si},\ O_s{}^i{}_j\in\Omega^1(B_0), \qquad U_i,\ V^i{}_j\in\Omega^0(B_0).
\end{equation}
The fields conjugate to rotational connections, curvatures, and
curvature-source quantities are understood in the $\mathfrak{so}(3)$ sector;
thus $M_c$, $M_d$, $O_d$, $O_s$, and $V$ are skew in their internal indices.
The three stress-field systems coexist whenever the corresponding structural
components are present. Their simultaneous presence does not imply that the
three geometric processes are identical. The compatible component describes
compatible deformation, whereas the distributed-incompatibility and
defect-source components describe evolution of internal structure.
Nevertheless, deformation and structural evolution may take place at the
same time. Their coupling enters through the cumulative geometric dependence
of the fields and, after the constitutive postulates are specified, through
the constitutive response of the material.

\subsection{Variational Formulation and Constitutive Postulates}

We use the combined geometric state introduced in the preceding sections, 
\begin{equation}
\mathcal{Z}=(\mathcal{Z}_c,\mathcal{Z}_d,\mathcal{Z}_s),
\end{equation}
with 
\begin{equation}
\mathcal{Z}_c=(e_c,\omega_c),\qquad \mathcal{Z}_d=(e_d,\omega_d;T_d,%
\Omega_d),\qquad \mathcal{Z}_s=(T_s,\Omega_s;S,R).
\end{equation}
This notation is recalled here only for use in the variational formulation;
it does not introduce a new state description.

For a potential-based material we postulate a single total stored-energy
three-form 
\begin{equation}
W=W(\mathcal{Z}) =W(e_c,\omega_c,e_d,\omega_d,T_d,\Omega_d,T_s,\Omega_s,S,R).
\end{equation}
The displayed dependence does not imply that all arguments are independently
variable. The geometric dependencies established above remain in force: $T_d$
and $\Omega_d$ are generated by $(e_d,\omega_d)$, while $S$ and $R$ are
generated from $(T_s,\Omega_s)$ on the retained defect geometry.

The stress fields postulated in the preceding subsection are assigned
constitutively by differentiation of the single total stored-energy
three-form with respect to the corresponding geometric arguments. With the
sign convention adopted in the present work, 
\begin{equation}
\Sigma_{ci}=-\frac{\partial W}{\partial e_c^i},\qquad M_c^j{}_{i}=-2\frac{%
\partial W}{\partial\omega_c^i{}_{j}},
\end{equation}
\begin{equation}
\Sigma_{di}=-\frac{\partial W}{\partial e_d^i},\qquad M_d^j{}_{i}=-2\frac{%
\partial W}{\partial\omega_d^i{}_{j}},
\end{equation}
\begin{equation}
H_{di}=\frac{\partial W}{\partial T_d^i},\qquad O_d^j{}_{i}=2\frac{\partial W%
}{\partial\Omega_d^i{}_{j}},
\end{equation}
\begin{equation}
H_{si}=\frac{\partial W}{\partial T_s^i},\qquad O_s^j{}_{i}=2\frac{\partial W%
}{\partial\Omega_s^i{}_{j}},
\end{equation}
\begin{equation}
U_i=\frac{\partial W}{\partial S^i},\qquad V^j{}_{i}=2\frac{\partial W}{%
\partial R^i{}_{j}}.
\end{equation}
These are constitutive partial derivatives of the total energy with respect
to the indicated geometric arguments. They do not make all arguments of $W$
independent kinematic variables. The geometric relations defining $T_d$, $%
\Omega_d$, $S$, and $R$ are imposed in the variation below. Thus the
constitutive dependence remains cumulative through the single function $W(%
\mathcal{Z})$, while the admissible variations retain the geometric
structure established in the preceding sections.

The variation is organized according to the independently variable geometric
fields and the derived fields generated by them. With the sign convention
adopted in the present work, the first variation of the single total
stored-energy three-form is 
\begin{align}
\delta W={}& -\delta e_{c}^{i}\wedge \Sigma _{ci}-\frac{1}{2}\delta \omega
_{c}^{i}{}_{j}\wedge M_{c}^{j}{}_{i}  \notag \\
& -\delta e_{d}^{i}\wedge \Sigma _{di}-\frac{1}{2}\delta \omega
_{d}^{i}{}_{j}\wedge M_{d}^{j}{}_{i}  \notag \\
& +\delta T_{d}^{i}\wedge H_{di}+\frac{1}{2}\delta \Omega
_{d}^{i}{}_{j}\wedge O_{d}^{j}{}_{i}  \notag \\
& +\delta T_s^{i}\wedge H_{si}+\frac{1}{2}\delta \Omega_s ^{i}{}_{j}\wedge
O_{s}^{j}{}_{i}  \notag \\
& +U_{i}\,\delta S^{i}+\frac{1}{2}V^{j}{}_{i}\,\delta R^{i}{}_{j}.
\end{align}%
The single variation $\delta W$ represents the energetic response of the
material body and permits direct constitutive coupling among all structural
components.

All independent and derived variations above are material variations on $B_0$%
. The current placement $x=\phi_c(X,\lambda)$ is therefore understood
through the material representation of the fields. At each fixed value of
the loading-path parameter $\lambda$, the cumulative variational problem is
evaluated on the corresponding instantaneous configuration. The virtual
variation $\delta$ is distinct from differentiation $d_{\lambda}$ along the
family of neighboring equilibrated states. Under the quasistatic
specialization stated above, no additional independent transport variation
is introduced into the Euler--Lagrange system. Coupling among compatible
deformation and structural evolution remains contained in the single total
energy $W(\mathcal{Z})$ and in the geometric dependencies used in the
variation.

The differential-form degrees are consistent in every term of $\delta W$:
one-form variations are paired with two-form stresses, two-form variations
with one-form stresses, and three-form variations with zero-form stresses.
Each term is therefore a three-form on $B_0$.

\subsection{Euler--Lagrange Equations of the Cumulative System}

The proposed framework is governed by one variational principle rather than
by a hierarchy of separately corrected field systems. Let 
\begin{equation}
\mathcal{A}[\mathcal{Z}] = \int_{B_0}W(\mathcal{Z})-\mathcal{W}_{\mathrm{ext}%
}
\end{equation}
be the total potential functional, where $W$ is the single total
stored-energy three-form introduced above. Equilibrium requires 
\begin{equation}
\delta\mathcal{A}=0
\end{equation}
for all admissible variations.

For the compatible component the admissible variations are generated by
material translation and rotation, 
\begin{equation}
\delta e_c^i = D_c\xi^i-\zeta^i{}_{j}e_c^j, \qquad \delta\omega_c^{ij} =
D_c\zeta^{ij}, \qquad \zeta^{ij}=-\zeta^{ji}.
\end{equation}

These variations are restricted by the compatibility of the compatible
component, $T_c=0$ and $\Omega_c=0$, and therefore preserve its
representation as a compatible Cosserat material structure. No corresponding
compatibility constraint is imposed on the distributed-incompatibility
component; hence $e_d$ and $\omega_d$ are independently varied, while the
variations of $T_d$ and $\Omega_d$ are induced by their geometric
definitions.

Together with these, the independent geometric variations of the internal
structural components are 
\begin{equation}
\delta e_d,\qquad \delta\omega_d,\qquad \delta T_s,\qquad \delta\Omega_s,
\end{equation}
subject to the admissibility conditions of the corresponding structures.
Here $\delta$ denotes an admissible virtual variation at a fixed value of
the loading-path parameter $\lambda$ and must be distinguished from the
loading-path differential $d_{\lambda}$. The latter describes passage
between neighboring equilibrated combined states along the actual loading
path. The material-response strategy does not alter the geometric
independence of the structural fields or the form of their Euler--Lagrange
equations; rather, it governs their incremental participation in
continuation of the cumulative response. A retained structural component may
therefore remain present while its loading-path increment vanishes over a
portion of the response. All field equations are written in material
representation on $B_0$, and under the quasistatic specialization introduced
above no additional independent transport variation is required.

The remaining variations are generated by the geometric definitions. In
particular, 
\begin{equation}
\delta T_d^i = D_d\delta e_d^i + \delta\omega_d^i{}_{j}\wedge e_d^j,
\end{equation}
\begin{equation}
\delta\Omega_d^i{}_{j} = D_d\delta\omega_d^i{}_{j},
\end{equation}
while 
\begin{align}
\delta S^i ={}& D_d\delta T_s^i + \delta\omega_d^i{}_{k}\wedge T_s^k -
\delta\Omega_s^i{}_{j}\wedge e_d^j - \Omega_s^i{}_{j}\wedge\delta e_d^j,
\end{align}
and 
\begin{align}
\delta R^i{}_{j} ={}& D_d\delta\Omega_s^i{}_{j} +
\delta\omega_d^i{}_{k}\wedge\Omega_s^k{}_{j} -
\delta\omega_d^k{}_{j}\wedge\Omega_s^i{}_{k}.
\end{align}

Thus all independent and derived variations are specified before the field
equations are obtained. Substitution of these variations into the first
variation of the total stored energy, followed by covariant integration by
parts, yields the Euler--Lagrange equations of the coupled system.

For the compatible component, the bulk equations are 
\begin{equation}
D_c\Sigma_{ci}=0,
\end{equation}
\begin{equation}
D_cM_c^{ij}+2e_c^{[i}\wedge\Sigma_c^{j]}=0.
\end{equation}
Because $\Sigma_c$ and $M_c$ are derivatives of the total energy $W(\mathcal{%
Z})$, these equations already contain the constitutive influence of all
structural components. No separate component-reaction fields are required.

For the distributed-incompatibility component, substitution of the
variations displayed above, including the $e_d$- and $\omega_d$-dependence
of $S$ and $R$, gives the Euler--Lagrange equations directly from the same
variation. Before interpreting these equations, it is important to
distinguish the coframe-conjugate stresses appearing at Levels I and II.
Although both are denoted by $\Sigma$, with the structural component
indicated by a subscript, they are conjugate to different geometric
variables of the combined state. At Level I, 
\begin{equation}
\Sigma_{ci}=-\frac{\partial W}{\partial e_c^i}
\end{equation}
is conjugate to the compatible coframe $e_c^i$ and describes the energetic
response associated with variation of the compatible Cosserat deformation.
At Level II, 
\begin{equation}
\Sigma_{di}=-\frac{\partial W}{\partial e_d^i}
\end{equation}
is instead conjugate to the incompatible coframe $e_d^i$ and measures the
energetic response to variation of the distributed-incompatibility geometry.
Thus $\Sigma_d$ is not merely the Level-II notation for the macroscopic
force stress $\Sigma_c$. The distinction is variational: 
\begin{equation}
\Sigma_{ci}\longleftrightarrow\delta e_c^i,\qquad
\Sigma_{di}\longleftrightarrow\delta e_d^i.
\end{equation}
Because $e_c$ and $e_d$ are distinct components of the combined state, their
conjugate stresses are correspondingly distinct, although both are obtained
from the same cumulative stored energy $W(\mathcal{Z})$. The common energy
therefore permits constitutive coupling of the levels without identifying
their variations or their conjugate stresses.

The Level-II coframe equation is 
\begin{equation}
D_dH_{di}-\Sigma_{di}-U_k\Omega_s^k{}_{i}=0.
\end{equation}
The first two terms constitute the intrinsic Level-II coframe balance. Here $%
H_{di}=\partial W/\partial T_d^i$ is conjugate to the distributed torsion $%
T_d^i$. Since 
\begin{equation}
T_d^i=D_de_d^i,
\end{equation}
the coframe part of its variation is $D_d\delta e_d^i$, and covariant
integration by parts produces $D_dH_{di}$. Hence $D_dH_{di}-\Sigma_{di}$ is
the intrinsic Level-II balance that would remain in the absence of Level-III
source coupling. This differs structurally from the compatible Level-I
equation $D_c\Sigma_{ci}=0$: the incompatible coframe enters through its
torsion and therefore carries the additional torsion-conjugate stress $H_d$.

The term $-U_k\Omega_s^k{}_{i}$ is the feedback of the Level-III source
structure on the Level-II coframe equation. It follows from the recursive
relation 
\begin{equation}
S^k=D_dT_s^k-\Omega_s^k{}_{j}\wedge e_d^j.
\end{equation}
Variation with respect to $e_d^i$ contains the contribution 
\begin{equation}
\delta S^k\supset-\Omega_s^k{}_{i}\wedge\delta e_d^i.
\end{equation}
When paired with $U_k=\partial W/\partial S^k$, this dependence generates
the source-feedback term in the Level-II coframe equation. It is therefore
not an independently postulated interaction force, but a consequence of the
recursive source geometry and the single cumulative energy. Schematically,
the equation has the structure 
\begin{equation}
\left(D_dH_{di}-\Sigma_{di}\right)-U_k\Omega_s^k{}_{i}=0,
\end{equation}
namely an intrinsic Level-II coframe balance supplemented by Level-III
feedback.

The Level-II connection equation can be written directly in antisymmetric
form as 
\begin{equation}
M_d{}^j{}_{i}-D_dO_d{}^j{}_{i}-2e_d^{[j}\wedge H_d{}_{i]}
-2U_{[i}T_s^{j]}-V^k{}_{i}\Omega_s^j{}_{k} +V^j{}_{k}\Omega_s^k{}_{i}=0.
\end{equation}
The first two terms are the direct connection and curvature contributions at
Level II. The field $M_d{}^j{}_{i}$ is conjugate to the incompatible
rotational connection $\omega_d^i{}_{j}$, whereas $O_d{}^j{}_{i}$ is
conjugate to the distributed curvature $\Omega_d^i{}_{j}$. Since 
\begin{equation}
\delta\Omega_d^i{}_{j}=D_d\delta\omega_d^i{}_{j},
\end{equation}
covariant integration by parts produces the derivative term $D_dO_d$. The
overall sign of the equation has been chosen so that the direct connection
stress $M_d$ appears with positive sign.

The antisymmetric coframe--torsion term is 
\begin{equation}
2e_d^{[j}\wedge H_d{}_{i]} =e_d^j\wedge H_{di}-e_d^i\wedge H_{dj}.
\end{equation}
It arises from the connection-dependent part of the torsion variation, 
\begin{equation}
\delta T_d^i=D_d\delta e_d^i+\delta\omega_d^i{}_{j}\wedge e_d^j.
\end{equation}
Because $\delta\omega_d^{ij}$ is antisymmetric, only the antisymmetric part
of $e_d^j\wedge H_{di}$ contributes. Thus $2e_d^{[j}\wedge H_d{}_{i]}$ is
the moment generated by the torsion-conjugate stress relative to the
incompatible coframe. It is the geometric analogue, at the
distributed-incompatibility level, of the coframe--stress contribution in a
Cosserat rotational balance.

The remaining terms express feedback of the Level-III source structure on
the Level-II connection equation. The source relation $S^i=D_dT_s^i-%
\Omega_s^i{}_{j}\wedge e_d^j$ depends on $\omega_d$ through $D_dT_s^i$. Its
connection variation therefore produces the antisymmetric term 
\begin{equation}
2U_{[i}T_s^{j]}=U_iT_s^j-U_jT_s^i,
\end{equation}
which represents the moment-type feedback associated with the torsional part
of the Level-III source structure. Likewise, 
\begin{equation}
R^i{}_{j}=D_d\Omega_s^i{}_{j}
\end{equation}
depends on the Level-II connection through the covariant derivative. Its
variation produces the curvature-source feedback 
\begin{equation}
-V^k{}_{i}\Omega_s^j{}_{k}+V^j{}_{k}\Omega_s^k{}_{i},
\end{equation}
where $V^j{}_{i}=2\partial W/\partial R^i{}_{j}$.

Consequently, the Level-II connection equation consists of an intrinsic
rotational balance, 
\begin{equation}
M_d{}^j{}_{i}-D_dO_d{}^j{}_{i}-2e_d^{[j}\wedge H_d{}_{i]},
\end{equation}
supplemented by the Level-III feedback generated through $S$ and $R$. The
coupling terms are not separately postulated interaction stresses. They
arise automatically because the Level-III source fields are defined relative
to the retained Level-II geometry. This makes explicit a central feature of
the cumulative construction: the structural variables remain geometrically
distinct, while the recursive definitions transmit the higher-level
energetic response back into the lower-level Euler--Lagrange equations.

For the defect-source component, $T_s^i$ and $\Omega_s^i{}_{j}$ are
independent. The corresponding source fields 
\begin{equation}
S^i=D_dT_s^i-\Omega_s^i{}_{j}\wedge e_d^j,\qquad
R^i{}_{j}=D_d\Omega_s^i{}_{j}
\end{equation}
are derived relative to the retained Level-II geometry $(e_d,\omega_d)$.
Consequently, the Level-III Euler--Lagrange equations are not independent of
the Level-II geometric background, even though $T_s$ and $\Omega_s$ are
independent Level-III variables.

Variation with respect to $T_s^i$ gives 
\begin{equation}
\delta S^i=D_d\delta T_s^i
\end{equation}
when $e_d$, $\omega_d$, and $\Omega_s$ are held fixed in this part of the
variation. The terms in $\delta W$ containing $\delta T_s^i$ are therefore 
\begin{equation}
\delta T_s^i\wedge H_{si}+U_iD_d\delta T_s^i.
\end{equation}
Covariant integration by parts transfers $D_d$ from $\delta T_s^i$ to $U_i$
and gives the bulk equation 
\begin{equation}
H_{si}-D_dU_i=0.
\end{equation}
Here $H_{si}=\partial W/\partial T_s^i$ is the one-form stress conjugate to
the independent Level-III torsional defect field $T_s^i$, whereas $%
U_i=\partial W/\partial S^i$ is the zero-form stress conjugate to its
source/nonclosure $S^i$. Thus the equation balances the direct energetic
response to $T_s$ against the covariant variation of the source-conjugate
quantity $U$. This should be distinguished from the Level-II quantity $H_{di}
$, which is conjugate to the derived torsion $T_d^i=D_de_d^i$. The symbols $%
H_d$ and $H_s$ therefore refer to different geometric arguments of the same
total energy $W(\mathcal{Z})$.

Variation with respect to $\Omega_s^i{}_{j}$ gives 
\begin{equation}
\delta S^i=-\delta\Omega_s^i{}_{j}\wedge e_d^j, \qquad \delta
R^i{}_{j}=D_d\delta\Omega_s^i{}_{j}.
\end{equation}
The first term expresses the fact that the rotational Level-III defect field
enters the translational source $S$ through the retained Level-II coframe,
while the second expresses the covariant source relation for $R$.
Substitution into $\delta W$, followed by covariant integration by parts and
use of the antisymmetry of $\delta\Omega_s^{ij}$, yields 
\begin{equation}
O_s^{ij}-D_dV^{ij}-2e_d^{[i}U^{j]}=0.
\end{equation}
The first two terms have the same variational pattern as the preceding
Level-III equation: $O_s^{ij}$ is conjugate to $\Omega_s^{ij}$, while $%
D_dV^{ij}$ is generated by the derivative term in $R^{ij}=D_d\Omega_s^{ij}$.
The last term is different. It arises from the explicit occurrence of $%
\Omega_s$ in $S^i=D_dT_s^i-\Omega_s^i{}_{j}\wedge e_d^j$. Because the
rotational variation is antisymmetric, only the antisymmetric part of the
product $e_d^iU^j$ contributes, namely 
\begin{equation}
2e_d^{[i}U^{j]}=e_d^iU^j-e_d^jU^i.
\end{equation}
It may therefore be interpreted as the moment-type coupling of the
source-conjugate field $U$ with the retained incompatible coframe $e_d$.
Unlike the Level-II term $2e_d^{[j}\wedge H_{d\,i]}$, which originates from
the dependence of the derived torsion $T_d=D_de_d$ on the Level-II
connection, the present term originates from the dependence of the Level-III
source $S$ on the independent curvature-type field $\Omega_s$. The algebraic
similarity of the two terms thus reflects the common rotational
antisymmetry, while their geometric origins and conjugate quantities are
distinct.

These equations make the recursive character of the cumulative framework
explicit. Level III introduces the independent defect fields $(T_s,\Omega_s)$%
, but their source/nonclosure measures $(S,R)$ are formed with the Level-II
covariant derivative $D_d$ and the retained Level-II coframe $e_d$. Hence
the Level-III equilibrium equations are equations for independent
higher-level fields represented on the accumulated lower-level geometry.
Conversely, as shown in the Level-II equations above, variation of that
lower-level geometry produces feedback terms containing the Level-III
stresses. The recursion is therefore variational in both directions without
identifying the geometric variables of the different levels.

Together with the geometric definitions of $T_d$, $\Omega_d$, $S$, and $R$
stated above, these equations form the bulk Euler--Lagrange system of the
cumulative variational problem.

Thus all component interactions arise automatically when the derived
geometric variations are substituted into the variation of the single total
energy. Terms generated in this way may be interpreted as variational
reactions of the coupled structural components, but they are not introduced
as additional fields and require no separate reaction notation.

\subsection{Boundary Conditions}

The physical boundary conditions are prescribed for the material body as a
whole and not separately for the individual structural contributions. This
common-boundary-data principle is a direct consequence of the cumulative
construction: changing the relative participation of $\mathcal{Z}_c$, $%
\mathcal{Z}_d$, and $\mathcal{Z}_s$ does not create a new physical specimen
or a new externally prescribed boundary-value problem.

Let $\mathcal{B}$ denote the physical boundary data. In a conventional
Cosserat representation these consist of complementary kinematic and
mechanical conditions, for example displacement or traction and rotation or
couple traction, 
\begin{equation}
u=\bar u\ \hbox{on }\Gamma_u,\qquad t=\bar t\ \hbox{on }\Gamma_t,
\end{equation}
\begin{equation}
\varphi=\bar\varphi\ \hbox{on }\Gamma_\varphi,\qquad m=\bar m\ \hbox{on }%
\Gamma_m,
\end{equation}
with the corresponding intrinsic formulation in terms of the coframe,
rotational connection, and their conjugate force and moment quantities.
These externally prescribed data remain of the same type whether the
response is predominantly compatible, contains substantial mesoelastic
incompatibility, or involves the complete defect-source structure.

The additional distributed-incompatibility and defect-source fields are
determined internally by their governing equations, constitutive relations,
geometric constraints, and the natural conditions generated by the
cumulative variational principle. Such natural conditions must be
distinguished from additional independently controlled physical loading
data. Thus the physical boundary data may be common to all combined states
even though integration by parts in the higher-level variations produces
boundary terms associated with the internal fields.

Accordingly, the common physical boundary and initial data are the
externally prescribed data of the theory; mathematical closure is obtained
only when they are taken together with the recursive field equations,
constitutive relations, geometric identities, and the natural variational
conditions of the internal fields. Closure in this sense does not imply
uniqueness: multiple equilibria or bifurcating branches may occur and may be
physically relevant.

The same common boundary specification is already implicit in the stability
calculations below. Perturbations are taken about states belonging to one
physical boundary-value problem and satisfy the corresponding homogeneous
restrictions. Stability therefore provides only information about a
particular combined-state equilibrium under those data; it does not define the
boundary conditions or the recursive structure of the theory.

\section{Illustrative Consequences of the Cumulative Framework}

The following illustrations are different consequences of one common reduced
cumulative model. The constitutive energy is therefore introduced once below
and is not redefined for each illustration. The purpose of the reduction is
not calibration of a particular material or definition of a universal
constitutive law, but a transparent demonstration of the capability of the
cumulative framework to represent characteristic mechanical features of solid
deformation. In this reduced setting, the cumulative material strategy
governs which structural components participate in the response, while the
constitutive energy determines how that participation appears mechanically.
Successive structural participation can therefore be related to changes of
tangent stiffness, unloading--reloading asymmetry, residual deformation,
hysteresis-type response, and reverse-loading effects.

The homogeneous model is quasistatic. Neighboring equilibrium states are
indexed by the loading-path parameter $\lambda$; no physical-time evolution
law is assumed unless it is introduced explicitly for a later kinetic
specialization. Compatible deformation may continue while the structural
variables evolve. Consistently with the quasistatic specialization above,
its increment during an active structural-evolution stage is treated as
small relative to the accompanying structural increment. The successive
illustrations examine loading and unloading, dependence on the initial
combined state, reverse loading, cyclic response, changes of tangent
response, and finally a spatial extension of the same model.

The reduced variables are

\begin{equation}
e_c,\qquad T_d,\qquad S.
\end{equation}

Here $e_c$ is a scalar compatible strain, $T_d$ is the material amplitude of
a selected translational defect mode, and $S$ is the material amplitude of a
selected defect-source mode. The rotational channel is suppressed in this
reduction,

\begin{equation}
\Omega_d=0,\qquad R=0.
\end{equation}

In the complete three-dimensional theory, $T_d=D_de_d$ is derived from the
defect coframe and connection, while the source component contains
independently variable defect fields together with their nonclosure
characteristics. The scalar variables used below are reduced internal
coordinates obtained after restricting the full field system; their use does
not change the variational status of the corresponding fields in the
three-dimensional theory.

All structural variables are formulated materially on $B_0$. When a spatial
representation is useful, the compatible deformation supplies the
corresponding spatial image; this is a change of representation and not an
additional transport law.

\paragraph{Common nonlinear cumulative potential}

For the bifurcation calculation we use the homogeneous reduced potential

\begin{equation}
\begin{split}
\Pi_{\lambda}(e_c,T_d,S)={}& \frac{1}{2}Ee_c^2-P(\lambda)e_c +\frac{1}{2}%
\left(a_T-2\alpha e_c-\gamma_2e_c^2\right)T_d^2 +\frac{b_T}{4}T_d^4 \\
&+\frac{1}{2}\left(\beta_S-\gamma_3T_d^2\right)S^2 +\frac{b_S}{4}S^4 .
\end{split}%
\end{equation}

Here $E,a_T,\beta_S,b_T,b_S,\alpha,\gamma_2,\gamma_3$ are positive material
parameters. The quartic structural part in $T_d$ and $S$ is bounded from
below when

\begin{equation}
b_Tb_S>\gamma_3^2 .
\end{equation}

Because the coupling $-\gamma_2e_c^2T_d^2/2$ is retained, the reduced
potential is used here as a local constitutive model over the deformation
range relevant to the bifurcation and unloading illustrations, rather than
as a globally coercive energy for arbitrarily large $e_c$.

The coefficient of $T_d^2$ decreases with compatible deformation through the
terms $-\alpha e_cT_d^2$ and $-\gamma_2e_c^2T_d^2/2$. The additional linear
dependence on $e_c$ does not destroy the stationarity of the perfect branch
because the first derivative with respect to $T_d$ still vanishes at $T_d=0$%
. It has a second purpose: once a nonzero defect structure is retained, it
can shift the zero-force compatible strain and therefore represent geometric
memory during ordinary unloading. The coupling $-\gamma_3T_d^2S^2/2$ permits
the defect structure to reduce the source stiffness.

\paragraph{Equilibrium equations}

Stationarity with respect to the three reduced variables gives

\begin{equation}
P(\lambda) = Ee_c-\left(\alpha+\gamma_2e_c\right)T_d^2 ,
\end{equation}

\begin{equation}
T_d\left[ a_T-2\alpha e_c-\gamma_2e_c^2+b_TT_d^2-\gamma_3S^2 \right]=0,
\end{equation}

and

\begin{equation}
S\left[ \beta_S-\gamma_3T_d^2+b_SS^2 \right]=0 .
\end{equation}

These equations admit, within one smooth constitutive potential, a
compatible branch, a compatible--defect branch, and a three-component
cumulative branch.

\paragraph{Cumulative tangent matrix}

The local incremental stability of a homogeneous equilibrium state is
governed by the Hessian of the total potential with respect to $(e_c,T_d,S)$,

\begin{equation}
K_{\mathrm{cum}}=\frac{\partial ^{2}\Pi }{\partial (e_{c},T_{d},S)^{2}}.
\end{equation}

Its explicit form is

\begin{equation}
K_{\mathrm{cum}}=%
\begin{pmatrix}
E-\gamma _{2}T_{d}^{2} & -2\left( \alpha +\gamma _{2}e_{c}\right) T_{d} & 0
\\ 
-2\left( \alpha +\gamma _{2}e_{c}\right) T_{d} & a_{T}-2\alpha e_{c}-\gamma
_{2}e_{c}^{2}+3b_{T}T_{d}^{2}-\gamma _{3}S^{2} & -2\gamma _{3}T_{d}S \\ 
0 & -2\gamma _{3}T_{d}S & \beta _{S}-\gamma _{3}T_{d}^{2}+3b_{S}S^{2}%
\end{pmatrix}%
.
\end{equation}

The matrix is symmetric, as required for the Hessian of a scalar potential.
For prescribed compatible strain, the structural tangent is the lower $%
2\times2$ block of $K_{\mathrm{cum}}$. Under dead-load control the complete $%
3\times3$ Hessian must remain positive definite. The full cumulative
tangent, the structural tangent, and the observable branch tangent therefore
play different roles.

\paragraph{Role of nonlinearity}

The nonlinear character of the reduced potential is essential to the
illustrations that follow. It is not introduced merely to obtain curved
force--deformation diagrams. The coupling terms make the stability and
equilibrium of each response depend on the cumulative structural state
already attained. In particular, the effective defect stiffness 
\begin{equation}
A_T(e_c)=a_T-2\alpha e_c-\gamma_2e_c^2
\end{equation}
depends on the compatible deformation. Increasing $e_c$ can therefore drive
the defect direction through a loss of stability and make the $Z_2$ response
accessible. Likewise, the effective source stiffness 
\begin{equation}
A_S(T_d)=\beta_S-\gamma_3T_d^2
\end{equation}
depends on the defect structure already developed, so that the $Z_3$
response may become accessible on the accumulated $Z_1+Z_2$ background.

The quartic terms 
\begin{equation}
\frac{b_T}{4}T_d^4,\qquad \frac{b_S}{4}S^4
\end{equation}
stabilize finite-amplitude branches after the corresponding quadratic
stiffness loses positivity. Thus the same nonlinear potential can support
the cumulative progression 
\begin{equation}
Z_1\longrightarrow Z_1+Z_2\longrightarrow Z_1+Z_2+Z_3 .
\end{equation}
The coupling of $e_c$ and $T_d$ also shifts the compatible equilibrium when
a defect structure is retained, because 
\begin{equation}
P(\lambda)=(E-\gamma_2T_d^2)e_c-\alpha T_d^2 .
\end{equation}
Consequently, residual deformation, residual stress, modified tangent
response, and asymmetric reverse response arise as related consequences of
the same nonlinear cumulative constitutive model rather than as separate
assumptions introduced for individual illustrations.

\subsection{Implementation of the Material Response Strategy in Loading and
Unloading}

This illustration demonstrates the implementation of the postulated material
response strategy 
\begin{equation}
Z_1\prec Z_2\prec Z_3
\end{equation}
during loading and unloading of an initially perfect body. The ordering is a
priority of material response to changes of externally affecting conditions;
it is not a prescribed temporal sequence of material states. The realized
path is determined by the current combined state, the constitutive energy,
the applied boundary conditions, and the stability or energetic
accessibility of the available responses.

Let $\lambda$ denote a monotonically increasing parameter used only to
parametrize traversal of the complete loading--unloading path. Thus $\lambda$
continues to increase after reversal of the external loading. If $\mathcal{E}%
(\lambda)$ denotes the stored internal energy along the realized path, then 
\begin{equation}
\frac{d\mathcal{E}}{d\lambda}>0
\end{equation}
characterizes energy accumulation, whereas 
\begin{equation}
\frac{d\mathcal{E}}{d\lambda}<0
\end{equation}
characterizes energy release.

\paragraph{Perfect branch and first bifurcation}

For an initially perfect body,

\begin{equation}
T_d=0,\qquad S=0,
\end{equation}

and

\begin{equation}
P(\lambda)=Ee_c .
\end{equation}

On this branch,

\begin{equation}
K_{\mathrm{cum}}^{(1)}=%
\begin{pmatrix}
E & 0 & 0 \\ 
0 & a_{T}-2\alpha e_{c}-\gamma _{2}e_{c}^{2} & 0 \\ 
0 & 0 & \beta _{S}%
\end{pmatrix}%
.
\end{equation}

Since $E>0$ and $\beta_S>0$, the first structural instability occurs when

\begin{equation}
a_T-2\alpha e_{c,2}-\gamma_2e_{c,2}^2=0 .
\end{equation}

The positive critical strain is

\begin{equation}
e_{c,2} = \frac{ -\alpha+\left(\alpha^2+\gamma_2a_T\right)^{1/2} }{\gamma_2},
\end{equation}

and

\begin{equation}
P_2=Ee_{c,2}.
\end{equation}

Thus the first critical load is a genuine loss of positive definiteness of
the cumulative second variation in the defect direction.

\paragraph{Compatible--defect branch}

Beyond the first bifurcation, the branch with $S=0$ and $T_d\neq0$ satisfies

\begin{equation}
T_d^2 = \frac{ 2\alpha e_c+\gamma_2e_c^2-a_T }{b_T}.
\end{equation}

The force on this branch is

\begin{equation}
P_{12}(e_c) = Ee_c - \left(\alpha+\gamma_2e_c\right)T_d^2 .
\end{equation}

The defect-direction tangent coefficient becomes

\begin{equation}
K_{TT}=2b_TT_d^2>0,
\end{equation}

whereas the source-direction coefficient is

\begin{equation}
K_{SS} = \beta_S-\gamma_3T_d^2 .
\end{equation}

Because $K_{TS}=0$ on $S=0$, the second structural bifurcation occurs when

\begin{equation}
T_{d,3}^2=\frac{\beta_S}{\gamma_3}.
\end{equation}

Consequently,

\begin{equation}
e_{c,3} = \frac{ -\alpha+ \left[ \alpha^2+ \gamma_2 \left(
a_T+b_T\beta_S/\gamma_3 \right) \right]^{1/2} }{\gamma_2},
\end{equation}

and

\begin{equation}
P_3 = Ee_{c,3} - \left(\alpha+\gamma_2e_{c,3}\right) \frac{\beta_S}{\gamma_3}%
.
\end{equation}

Under dead-load control the compatible--defect subblock must remain positive
definite up to this second bifurcation. Its determinant is

\begin{equation}
\Delta_{cT} = 2T_d^2 \left[ b_T\left(E-\gamma_2T_d^2\right) -
2\left(\alpha+\gamma_2e_c\right)^2 \right].
\end{equation}

Accordingly, a clean successive source bifurcation under dead loading
requires $\Delta_{cT}>0$ before $K_{SS}$ reaches zero. Under prescribed $e_c$%
, the structural block alone determines the structural instability.

\paragraph{Three-component cumulative branch}

For $T_d\neq0$ and $S\neq0$,

\begin{equation}
S^2 = \frac{\gamma_3T_d^2-\beta_S}{b_S}.
\end{equation}

Substitution into the defect equation gives

\begin{equation}
T_d^2 = \frac{ 2\alpha e_c+\gamma_2e_c^2-a_T-\gamma_3\beta_S/b_S }{
b_T-\gamma_3^2/b_S }.
\end{equation}

The denominator is positive when

\begin{equation}
b_Tb_S>\gamma_3^2 .
\end{equation}

The force on this branch is

\begin{equation}
P_{123}(e_c) = Ee_c - \left(\alpha+\gamma_2e_c\right)T_d(e_c)^2 .
\end{equation}

At $e_c=e_{c,3}$, these formulas give $T_d^2=\beta_S/\gamma_3$ and $S=0$, so
the three-component branch joins continuously to the compatible--defect
branch.

\paragraph{Branch tangent moduli}

The observable branch tangent differs from the entries of $K_{\mathrm{cum}}$
because the internal variables evolve along the equilibrium branch. On the
perfect branch,

\begin{equation}
E_{\mathrm{tan}}^{(1)}=E .
\end{equation}

On the compatible--defect branch,

\begin{equation}
E_{\mathrm{tan}}^{(12)} = E-\gamma_2T_d^2 - \frac{ 2\left(\alpha+\gamma_2e_c%
\right)^2 }{b_T}.
\end{equation}

Let

\begin{equation}
D_3=b_T-\frac{\gamma_3^2}{b_S}.
\end{equation}

On the three-component branch,

\begin{equation}
E_{\mathrm{tan}}^{(123)} = E-\gamma_2T_d^2 - \frac{ 2\left(\alpha+%
\gamma_2e_c\right)^2 }{D_3}.
\end{equation}

These relations separate the full cumulative Hessian, the structural
stability submatrix, and the directly observable slope $dP/de_c$.

\paragraph{Initial compatible response.}

For the virgin specialization

\begin{equation}
T_d=0,\qquad S=0,
\end{equation}

the force law is $P(\lambda)=Ee_c$. The first departure from this branch is
not introduced by an external threshold rule. It occurs when the defect
direction of the cumulative second variation loses positive definiteness.
The first critical strain therefore marks a genuine structural bifurcation
of the selected reduced energy.

\paragraph{Changing structural participation.}

After the first bifurcation, $T_d\neq0$ while $S=0$, so the compatible and
defect components participate simultaneously. A second bifurcation may occur
when the source-direction tangent loses stability, after which all three
components participate. This sequence is specific to the virgin initial
state. The illustration of a nonvirgin initial state shows that a body with
pre-existing defect or source structure need not pass through these branches
in this order.

\paragraph{Cumulative loading path}

For the initially perfect state and the present constitutive specialization,
the realized energy-accumulation path is therefore 
\begin{equation}
Z_1\longrightarrow Z_1+Z_2\longrightarrow Z_1+Z_2+Z_3 .
\end{equation}
This sequence is not an additional postulate. It is generated by the single
response priority $Z_1\prec Z_2\prec Z_3$ together with the selected initial
state, constitutive energy, boundary loading, and successive stability
limits.

\paragraph{Unloading to the initial zero-load boundary conditions}

Suppose that the loading path has produced a nonzero defect structure and,
possibly, a nonzero source structure. The externally affecting conditions
are then reversed and unloading begins. Unloading continues until the
external boundary conditions return to their initial zero-load values. The
central question is whether return of the external boundary conditions to
their initial values also returns the cumulative material state to its
initial state.

\paragraph{Response sequence along the loading parameter}

Let $\lambda$ increase monotonically along the complete loading--unloading
history, so that reversal of mechanical loading does not require reversal of 
$\lambda$. At the beginning of unloading let 
\begin{equation}
\mathcal{Z}_u=(Z_{1u},Z_{2u},Z_{3u}).
\end{equation}
The same response priority applies during unloading. Along the restricted
unloading path considered here, the compatible response acts first on the
retained cumulative background, schematically, 
\begin{equation}
(Z_{1u},Z_{2u},Z_{3u}) \longrightarrow (Z_1(\lambda),Z_{2u},Z_{3u}).
\end{equation}
The changing $Z_1$ therefore does not imply disappearance of the previously
acquired $Z_2$ or $Z_3$ structure. Thus during the final $Z_1$ portion one
may have 
\begin{equation}
\mathcal{Z}(\lambda)= \left(\mathcal{Z}_c(\lambda);\mathcal{Z}_{d,r};%
\mathcal{Z}_{s,r}\right).
\end{equation}
This is the simplest explicit illustration of the strategic priority $%
Z_1\prec Z_2\prec Z_3$ during both loading and unloading.

\paragraph{Loaded combined state}

Let the maximum state reached during loading be

\begin{equation}
\left( e_{c,\max},T_{d,\max},S_{\max} \right), \qquad T_{d,\max}\neq0 .
\end{equation}

If the maximum state lies on the three-component branch, then also $%
S_{\max}\neq0$. Ordinary unloading changes the prescribed mechanical loading 
$P(\lambda)$. It does not, by itself, prescribe the independent reverse
configurational loading that would be required to drive the structural
variables back through their previous path.

For the reduced retention problem we therefore consider the special
unloading path

\begin{equation}
T_d=T_{d,r},\qquad S=S_r,
\end{equation}

where $T_{d,r}$ and $S_r$ are the structural values retained from the
previous loading history. In the simplest case,

\begin{equation}
T_{d,r}=T_{d,\max}, \qquad S_r=S_{\max}.
\end{equation}

This is not a statement that the structural variables are universally
irreversible. It is the reduced representation of ordinary unloading when
the reverse generalized structural loading is not independently supplied.

\paragraph{Mechanical unloading at fixed retained structure}

During this unloading path the internal structural variables are fixed and
the mechanical equilibrium equation from the same cumulative potential is

\begin{equation}
P(\lambda) = Ee_c- \left(\alpha+\gamma_2e_c\right)T_{d,r}^2 .
\end{equation}

Equivalently,

\begin{equation}
P(\lambda) = \left( E-\gamma_2T_{d,r}^2 \right)e_c - \alpha T_{d,r}^2 .
\end{equation}

The unloading tangent is therefore

\begin{equation}
E_{\mathrm{unload}} = \frac{dP(\lambda)}{de_c} = E-\gamma_2T_{d,r}^2 .
\end{equation}

A mechanically stable unloading path requires

\begin{equation}
E-\gamma_2T_{d,r}^2>0 .
\end{equation}

\paragraph{Residual compatible strain}

At zero external force,

\begin{equation}
P(\lambda)=0,
\end{equation}

and hence

\begin{equation}
e_{c,r} = \frac{ \alpha T_{d,r}^2 }{ E-\gamma_2T_{d,r}^2 }.
\end{equation}

Thus

\begin{equation}
P(\lambda)=0,\qquad T_{d,r}\neq0 \quad\Longrightarrow\quad e_{c,r}\neq0
\end{equation}

whenever $\alpha>0$ and the unloading tangent remains positive.

The source amplitude $S_r$ does not appear explicitly in the final
one-dimensional force relation. It nevertheless affects the retained
combined state because it influences the loading branch on which $T_{d,r}$
was produced and because it remains part of the structural state. In the
full theory the analogous source fields also couple through the tensorial
and differential-form constitutive structure.

\paragraph{Meaning of the residual state}

The residual compatible strain is not introduced as an independent plastic
strain. It is the equilibrium response of the compatible component to the
retained defect structure. The final state after ordinary unloading is

\begin{equation}
\mathcal{Z}_r = \left( \mathcal{Z}_{c,r}; \mathcal{Z}_{d,r}; \mathcal{Z}%
_{s,r} \right),
\end{equation}

with

\begin{equation}
\mathcal{Z}_{d,r}\neq0
\end{equation}

and, when source structure was generated during loading,

\begin{equation}
\mathcal{Z}_{s,r}\neq0 .
\end{equation}

Ordinary mechanical unloading therefore need not generate the reverse
sequence

\begin{equation}
(\mathcal{Z}_c;\mathcal{Z}_d;\mathcal{Z}_s) \longrightarrow (\mathcal{Z}_c;%
\mathcal{Z}_d;0) \longrightarrow (\mathcal{Z}_c;0;0).
\end{equation}

Such a reverse structural path would require the corresponding reverse
generalized loading or an additional constitutive evolution law that permits
the structural variables to relax.

\paragraph{Energy interpretation}

With $T_d=T_{d,r}$ and $S=S_r$ fixed, the unloading part of the process is
mechanically reversible with respect to $e_c$. Along this restricted path,

\begin{equation}
dW = P(\lambda)\,de_c .
\end{equation}

Therefore the mechanical unloading segment itself need not be assigned an
independent dissipation mechanism. Any nonzero cyclic work in a complete
loading--unloading history must be interpreted together with the structural
evolution that occurred when $T_d$ and $S$ changed, or with the work of the
generalized forces required to reverse them.

This distinction is important for the cumulative framework. Macroscopic
irreversibility under ordinary loading does not require every internal
structural process to be fundamentally irreversible. A structural state may
be reversible with respect to its own generalized loading while remaining
retained when only the ordinary mechanical load is reversed.

\paragraph{Full-field interpretation}

In the three-dimensional Cosserat theory the retained defect structure is
represented by the material fields

\begin{equation}
e_d^i,\qquad \omega_d{}^i{}_j,
\end{equation}

with derived densities

\begin{equation}
T_d^i=D_de_d^i, \qquad \Omega_d{}^i{}_j = d\omega_d{}^i{}_j+
\omega_d{}^i{}_k\wedge\omega_d{}^k{}_j.
\end{equation}

The source structure is represented by the independently variable defect
fields and their source/nonclosure fields. Ordinary unloading changes the
compatible deformation through the physical boundary loading, but it does
not automatically prescribe the reverse configurational loading required to
erase the retained material structure. The scalar residual strain above is
the simplest reduced manifestation of that general statement.

\paragraph{Residual deformation and structural memory.}

Ordinary mechanical unloading does not necessarily supply the reverse
generalized loading required to restore the internal structural state. If a
nonzero defect amplitude is retained, the zero-force compatible strain is
shifted. Residual deformation is therefore interpreted as the compatible
equilibrium response to retained structure rather than as an independently
postulated plastic strain.

\paragraph{Residual combined state}

The preceding calculation shows that the same cumulative potential used for
the bifurcation calculation can also represent residual deformation without
reintroducing a linear term that destroys the perfect stationary branch. The
coupling $-\alpha e_cT_d^2$ leaves $T_d=0$ stationary before the first
bifurcation, while a retained nonzero $T_d$ shifts the zero-force
equilibrium after unloading. Hence the model separates two questions:
stability determines when a structural component can appear, whereas the
availability or absence of reverse generalized structural loading determines
whether that component is removed during ordinary unloading.

\paragraph{Constrained unloading and residual stress}

The zero-load case above considered a body with retained defect and
defect-source structure. The external force was removed and the compatible
strain adjusted to a nonzero residual value. The present example considers
the complementary experiment: after the same retained structural state has
been created, the compatible deformation is constrained during unloading.
The retained structure then appears as a residual mechanical force, or in a
continuum interpretation as residual stress.

\paragraph{Constrained retained state}

Let the retained structural amplitudes be

\begin{equation}
T_d=T_{d,r},\qquad S=S_r,
\end{equation}

and prescribe the compatible strain during unloading as

\begin{equation}
e_c=\bar e_c.
\end{equation}

The force relation of the same reduced cumulative model is

\begin{equation}
P_r = \left(E-\gamma_2T_{d,r}^2\right)\bar e_c -\alpha T_{d,r}^2 .
\end{equation}

Thus the retained defect structure generates a nonzero mechanical reaction
whenever the imposed compatible strain differs from the free residual value.

If the body is returned geometrically to its original compatible
configuration,

\begin{equation}
\bar e_c=0,
\end{equation}

the residual force becomes

\begin{equation}
P_r=-\alpha T_{d,r}^2 .
\end{equation}

For $\alpha>0$ and $T_{d,r}\neq0$, this reaction is nonzero. In the
one-dimensional reduction it is the residual-force analogue of residual
stress.

\paragraph{Relation to free residual deformation}

The zero-load unloading state satisfies

\begin{equation}
0 = \left(E-\gamma_2T_{d,r}^2\right)e_{c,r} -\alpha T_{d,r}^2,
\end{equation}

and therefore

\begin{equation}
e_{c,r} = \frac{\alpha T_{d,r}^2} {E-\gamma_2T_{d,r}^2}.
\end{equation}

Consequently, residual deformation and residual stress are two boundary
responses of the same retained combined structural state. Under free
unloading the body changes its compatible deformation until the external
force vanishes. Under constrained unloading the compatible deformation is
prescribed, and the same retained structure is balanced by a nonzero
reaction force.

\paragraph{Interpretation}

This distinction is important for the cumulative framework and gives a direct
observable consequence of the response strategy. During unloading the body
first changes its compatible response $Z_1$ on the retained $Z_2$ and, when
present, $Z_3$ background. Residual strain and residual stress do not
require two independent internal mechanisms. They are different observable
consequences of retained defect and defect-source structure under different
mechanical boundary conditions. In a three-dimensional formulation the
scalar reaction $P_r$ is replaced by the corresponding residual force-stress
and, where applicable, couple-stress fields required to satisfy the imposed
compatibility and boundary constraints.

\paragraph{Boundary-condition interpretation}

This comparison shows that the retained combined state has both a free and
a constrained macroscopic manifestation. Unloading to zero external load
produces a residual compatible deformation at zero external force, whereas
constrained return to the original compatible configuration produces a
nonzero residual reaction. The distinction follows from the same
constitutive force law and does not require an additional plastic-strain
variable.

\paragraph{Closed external loading cycle and non-closed material-state path}

The essential consequence of the complete loading--unloading illustration is
that the external boundary conditions may return to their initial zero-load
values without restoration of the initial cumulative material state: 
\begin{equation}
P(\lambda_{\mathrm{final}})=P(\lambda_{\mathrm{initial}})=0, \qquad \mathcal{%
Z}_{\mathrm{final}} \neq \mathcal{Z}_{\mathrm{initial}} .
\end{equation}
Residual deformation under zero external load and residual stress under a
constraint on compatible return are therefore two boundary-condition
manifestations of the same retained cumulative structure. The external
loading cycle may close while the cumulative material-state path does not.

\subsection{Influence of the Initial Combined State}

\paragraph{Purpose}

Examples 1 and 2 used a virgin perfect initial state in order to display the
special sequence from compatible deformation to defect participation and
then to source participation. The cumulative framework itself does not require
that sequence. The present example shows explicitly how the response changes
when defect or source structure is already present at the beginning of the
loading process.

\paragraph{General initial combined state}

Let the initial homogeneous state be

\begin{equation}
\left( e_c^0,T_d^0,S^0 \right),
\end{equation}

where one or both structural amplitudes may be nonzero. The same cumulative
potential used in Examples 1 and 2 is retained,

\begin{equation}
\begin{split}
\Pi_{\lambda}(e_c,T_d,S)={}& \frac{1}{2}Ee_c^2-P(\lambda)e_c +\frac{1}{2}%
\left(a_T-2\alpha e_c-\gamma_2e_c^2\right)T_d^2 +\frac{b_T}{4}T_d^4 \\
&+\frac{1}{2}\left(\beta_S-\gamma_3T_d^2\right)S^2 +\frac{b_S}{4}S^4 .
\end{split}%
\end{equation}

No new macroscopic branch, threshold variable, or history parameter is
introduced. The influence of the initial structure enters through the
initial point at which the same equilibrium equations and the same
cumulative tangent are evaluated.

\paragraph{Initial force and tangent response}

At the initial state, the physical force required for equilibrium is

\begin{equation}
P^0 = Ee_c^0 - \left( \alpha+\gamma_2e_c^0 \right) \left(T_d^0\right)^2 .
\end{equation}

The cumulative tangent evaluated on the initial structural background is

\begin{equation}
\mathbf{K}_{\mathrm{cum}}^{0}=%
\begin{pmatrix}
E-\gamma _{2}\left( T_{d}^{0}\right) ^{2} & -2\left( \alpha +\gamma
_{2}e_{c}^{0}\right) T_{d}^{0} & 0 \\ 
-2\left( \alpha +\gamma _{2}e_{c}^{0}\right) T_{d}^{0} & K_{TT}^{0} & 
K_{TS}^{0} \\ 
0 & K_{TS}^{0} & K_{SS}^{0}%
\end{pmatrix}%
,
\end{equation}

where

\begin{equation}
K_{TT}^0 = a_T-2\alpha e_c^0-\gamma_2\left(e_c^0\right)^2
+3b_T\left(T_d^0\right)^2 -\gamma_3\left(S^0\right)^2,
\end{equation}

\begin{equation}
K_{TS}^0 = -2\gamma_3T_d^0S^0,
\end{equation}

and

\begin{equation}
K_{SS}^0 = \beta_S-\gamma_3\left(T_d^0\right)^2 +3b_S\left(S^0\right)^2 .
\end{equation}

Thus a nonzero initial defect structure immediately couples the compatible
and defect increments, while simultaneous nonzero defect and source
structure also couples the two structural increments. In such a state the
three components participate from the beginning of the loading process.

\paragraph{Structural tangent for prescribed compatible strain}

When the compatible strain is the prescribed loading parameter, the
incremental structural response is governed by

\begin{equation}
\mathbf{K}_{\mathrm{str}}^{0}=%
\begin{pmatrix}
K_{TT}^{0} & K_{TS}^{0} \\ 
K_{TS}^{0} & K_{SS}^{0}%
\end{pmatrix}%
.
\end{equation}

Local structural stability requires

\begin{equation}
K_{TT}^0>0
\end{equation}

and

\begin{equation}
K_{TT}^0K_{SS}^0-\left(K_{TS}^0\right)^2>0.
\end{equation}

Accordingly, when $T_d^0\neq0$ and $S^0\neq0$, the relevant instability is
generally a coupled structural mode. It is no longer meaningful to interpret
the response as a compulsory first activation of defects followed by a
compulsory activation of sources.

\paragraph{Important special cases}

If

\begin{equation}
T_d^0=0,\qquad S^0=0,
\end{equation}

the perfect-state result of Example 1 is recovered, and the first loss of
stability is determined by

\begin{equation}
a_T-2\alpha e_c-\gamma_2e_c^2=0.
\end{equation}

If

\begin{equation}
T_d^0\neq0,\qquad S^0=0,
\end{equation}

the defect component participates from the initial state. The source
direction remains uncoupled to first order and its tangent coefficient is

\begin{equation}
K_{SS} = \beta_S-\gamma_3T_d^2 .
\end{equation}

A later source instability may therefore still occur, but it occurs from an
already two-component initial background rather than from a virgin material.

If

\begin{equation}
T_d^0\neq0,\qquad S^0\neq0,
\end{equation}

all three components are present initially. Their relative participation
changes continuously with loading, and the stability limit is determined by
the smallest eigenvalue of the full cumulative tangent or, under prescribed $%
e_c$, by positive definiteness of the structural tangent, equivalently by
$K_{TT}^{0}>0$ together with $\det\mathbf{K}_{\mathrm{str}}^{0}>0$.

\paragraph{Interpretation}

This problem makes precise an important qualification of the cumulative
theory. The sequence

\begin{equation}
(\mathcal{Z}_{c};0;0)\longrightarrow (\mathcal{Z}_{c};\mathcal{Z}%
_{d};0)\longrightarrow (\mathcal{Z}_{c};\mathcal{Z}_{d};\mathcal{Z}_{s})
\end{equation}

is a special consequence of the perfect initial state used in Example 1. For
a general initial state,

\begin{equation}
\mathcal{Z}(0)=\left( \mathcal{Z}_{c}^{0};\mathcal{Z}_{d}^{0};\mathcal{Z}%
_{s}^{0}\right) ,
\end{equation}

with $\mathcal{Z}_{d}^{0}$ and $\mathcal{Z}_{s}^{0}$ not necessarily zero, the cumulative framework describes
simultaneous coupled participation of all components that are already
present. External loading changes their relative contributions; it does not
impose a universal temporal hierarchy.

\paragraph{Interpretive consequence}

The initial structural state enters the response through the same cumulative
energy, equilibrium equations, and tangent matrix used for the virgin case.
Pre-existing structure changes the initial force, the incremental stiffness,
and the stability conditions without requiring an additional macroscopic
constitutive branch. This provides a direct representation of structural
history in the combined state itself.

\subsection{Reverse Loading and Bauschinger-Type Response}

\paragraph{Purpose}

The purpose of this example is to compare reverse mechanical loading from a
retained combined state with loading from the original perfect state. The
reverse experiment does not begin from virgin material, but from the
residual state produced by the preceding loading--unloading history. The
retained structural variables therefore modify both the mechanical response
and the local structural stability problem. The example is intended to
demonstrate a Bauschinger-type capability of the cumulative model, namely
directional memory produced by retained structure, rather than a
quantitatively calibrated Bauschinger law.

\paragraph{Virgin response in the opposite direction}

For the perfect state,

\begin{equation}
T_d=0,\qquad S=0,
\end{equation}

the force law is

\begin{equation}
P(\lambda)=Ee_c.
\end{equation}

The defect-direction tangent coefficient is

\begin{equation}
K_{TT}=a_T-2\alpha e_c-\gamma_2e_c^2.
\end{equation}

Hence the two roots of the perfect-branch neutral condition are

\begin{equation}
e_c^{\pm}=\frac{-\alpha\pm\left(\alpha^2+\gamma_2a_T\right)^{1/2}}{\gamma_2},
\end{equation}

with corresponding virgin critical loads

\begin{equation}
P_2^{\pm}=Ee_c^{\pm}.
\end{equation}

For the numerical parameters used above,

\begin{equation}
e_c^+=0.10000,\qquad e_c^-=-0.12000,
\end{equation}

and therefore

\begin{equation}
P_2^+=10.0000,\qquad P_2^-=-12.0000.
\end{equation}

The unequal magnitudes arise from the constitutive coupling $-\alpha e_cT_d^2
$ and are therefore properties of the selected reduced material model, not
universal symmetry properties of the cumulative framework.

\paragraph{Reverse loading from a retained combined state}

Let ordinary unloading from a previous positive loading history leave

\begin{equation}
T_d=T_{d,r},\qquad S=S_r.
\end{equation}

As long as this retained structural state does not change, reverse
mechanical loading follows the same constitutive force relation as the
preceding unloading segment,

\begin{equation}
P(\lambda)=\left(E-\gamma_2T_{d,r}^2\right)e_c-\alpha T_{d,r}^2.
\end{equation}

The corresponding mechanical tangent is

\begin{equation}
E_r=\frac{dP}{de_c}=E-\gamma_2T_{d,r}^2.
\end{equation}

Thus the retained defect structure changes both the tangent and the
intercept of the reverse mechanical path. At $e_c=0$,

\begin{equation}
P(\lambda)=-\alpha T_{d,r}^2,
\end{equation}

whereas the virgin state gives $P=0$. Conversely, the retained-state path
crosses the zero-force axis at

\begin{equation}
e_{c,r}=\frac{\alpha T_{d,r}^2}{E-\gamma_2T_{d,r}^2}.
\end{equation}

This shift of the reverse path is a direct mechanical manifestation of the
retained combined state. No separate macroscopic back-stress variable is
introduced.

\paragraph{Local structural stability on the retained background}

If the retained structure is subsequently allowed to evolve, the local
structural tangent must be evaluated on the retained state rather than on
the virgin state. At $(e_c,T_{d,r},S_r)$,

\begin{equation}
K_{TT}^{r}=a_T-2\alpha e_c-\gamma_2e_c^2+3b_TT_{d,r}^2-\gamma_3S_r^2,
\end{equation}

\begin{equation}
K_{TS}^{r}=-2\gamma_3T_{d,r}S_r,
\end{equation}

and

\begin{equation}
K_{SS}^{r}=\beta_S-\gamma_3T_{d,r}^2+3b_SS_r^2.
\end{equation}

For $K_{SS}^{r}>0$, a local structural neutral point of the retained branch
is determined by

\begin{equation}
K_{TT}^{r}K_{SS}^{r}-\left(K_{TS}^{r}\right)^2=0,
\end{equation}

or equivalently

\begin{equation}
K_{TT}^{r}=\frac{\left(K_{TS}^{r}\right)^2}{K_{SS}^{r}}.
\end{equation}

This condition is generally different from the virgin condition $K_{TT}=0$.
When both $T_{d,r}$ and $S_r$ are nonzero, the incipient structural mode is
coupled. If a solution $e_c=e_{c,r}^{\ast}$ of this neutral condition is
reached while the retained branch remains admissible, the corresponding
mechanical load is

\begin{equation}
P_r^{\ast}=\left(E-\gamma_2T_{d,r}^2\right)e_{c,r}^{\ast}-\alpha T_{d,r}^2.
\end{equation}

The comparison of $P_r^{\ast}$ with the virgin value $P_2^-$ measures the
change of the local reverse stability limit generated by the retained
combined state. Its sign and magnitude are constitutive and need not
correspond to a universal reduction of the reverse critical load.

\paragraph{What the reduced model does and does not predict}

The conservative stored energy determines equilibrium and local stability,
but it does not by itself prescribe the kinetic rule by which a retained,
possibly metastable, structure is released during reverse loading. Therefore 
$P_r^{\ast}$ is a candidate local neutral load of the retained equilibrium
branch, not a universally determined reverse reactivation load. A unique
reactivation point requires an additional constitutive evolution or
dissipation law specifying which admissible structural path is followed.

This distinction is important for the present example. The cumulative model
already predicts that prior structural evolution changes the reverse
mechanical tangent, zero-force strain, and local stability condition. These
are Bauschinger-type manifestations of structural memory. A quantitative
Bauschinger prediction for a particular solid would additionally require
material calibration and an evolution law; neither is imposed here merely to
manufacture a desired threshold.

\paragraph{Interpretive consequence}

Virgin reverse loading begins from

\begin{equation}
\left(\mathcal{Z}_{c};0;0\right),
\end{equation}

whereas reverse loading after prior structural evolution begins from

\begin{equation}
\left(\mathcal{Z}_{c,r};\mathcal{Z}_{d,r};\mathcal{Z}_{s,r}\right).
\end{equation}

The two histories therefore start from different points of the same
combined-state space. The cumulative strategy organizes which structural
components may participate during reversal, while the constitutive energy
determines how the retained state changes the observable mechanical response
and its local stability. In this sense the Bauschinger-type response is not
appended as a separate macroscopic rule; it arises as a capability of the
same combined-state description used for loading, unloading, and residual
response. Here ``Bauschinger-type'' denotes directional memory manifested by
the shifted reverse mechanical path; a quantitative reverse reactivation
threshold requires an additional structural evolution law.

\subsection{Parametric Cyclic Loading and Cumulative Hysteresis}

\paragraph{Purpose}

The preceding examples show that the response of the body is determined not
only by the instantaneous value of an externally controlled loading condition
but also by the structural components retained during the preceding loading
path. This provides a natural setting for describing cyclic response without
introducing an additional time-dependent evolution law.

Let $\lambda$ denote a parameter describing the externally controlled loading
path, and write the combined material state as

\begin{equation}
\mathcal{Z}(\lambda)=\left(\mathcal{Z}_{c}(\lambda);\mathcal{Z}_{d}(\lambda);%
\mathcal{Z}_{s}(\lambda)\right).
\end{equation}

For the reduced model this becomes

\begin{equation}
\mathcal{Z}(\lambda)=\left(e_c(\lambda),T_d(\lambda),S(\lambda)\right).
\end{equation}

The parameter $\lambda$ labels the loading path and is not identified with
physical time. It may represent a prescribed displacement, compatible strain,
load amplitude, or another externally controlled condition.

\paragraph{Parametric loading and reversal}

Consider a cyclic variation of the external loading condition,

\begin{equation}
\lambda_0\longrightarrow\lambda_{\max}\longrightarrow\lambda_{\min}%
\longrightarrow\lambda_0.
\end{equation}

During the first loading branch, an initially compatible state may undergo
the successive structural activations described in Example 1. At the maximum
loading condition the combined state may therefore be written

\begin{equation}
\mathcal{Z}_{\max}=\left(e_{c,\max},T_{d,r},S_r\right),
\end{equation}

where $T_{d,r}$ and $S_r$ denote structural components retained upon reversal.
The essential point is that reversal of the loading parameter does not imply
reversal of the complete combined state. Thus, at the same compatible strain,

\begin{equation}
\mathcal{Z}_{\mathrm{load}}(e_c)\neq
\mathcal{Z}_{\mathrm{unload}}(e_c),
\end{equation}

and, in general,

\begin{equation}
T_d^{\mathrm{load}}(e_c)\neq T_d^{\mathrm{unload}}(e_c),\qquad
S^{\mathrm{load}}(e_c)\neq S^{\mathrm{unload}}(e_c).
\end{equation}

\paragraph{Macroscopic hysteresis}

The observable force is

\begin{equation}
P(\lambda)=Ee_c-\left(\alpha+\gamma_2e_c\right)T_d^2.
\end{equation}

Consequently, different combined structural states at the same value of
$e_c$ may give different macroscopic forces,

\begin{equation}
P_{\mathrm{load}}(e_c)\neq P_{\mathrm{unload}}(e_c).
\end{equation}

The loading and unloading branches therefore need not retrace one another in
the $(e_c,P)$ plane.

\begin{figure}[htbp]
\centering
\includegraphics[width=0.72\textwidth]{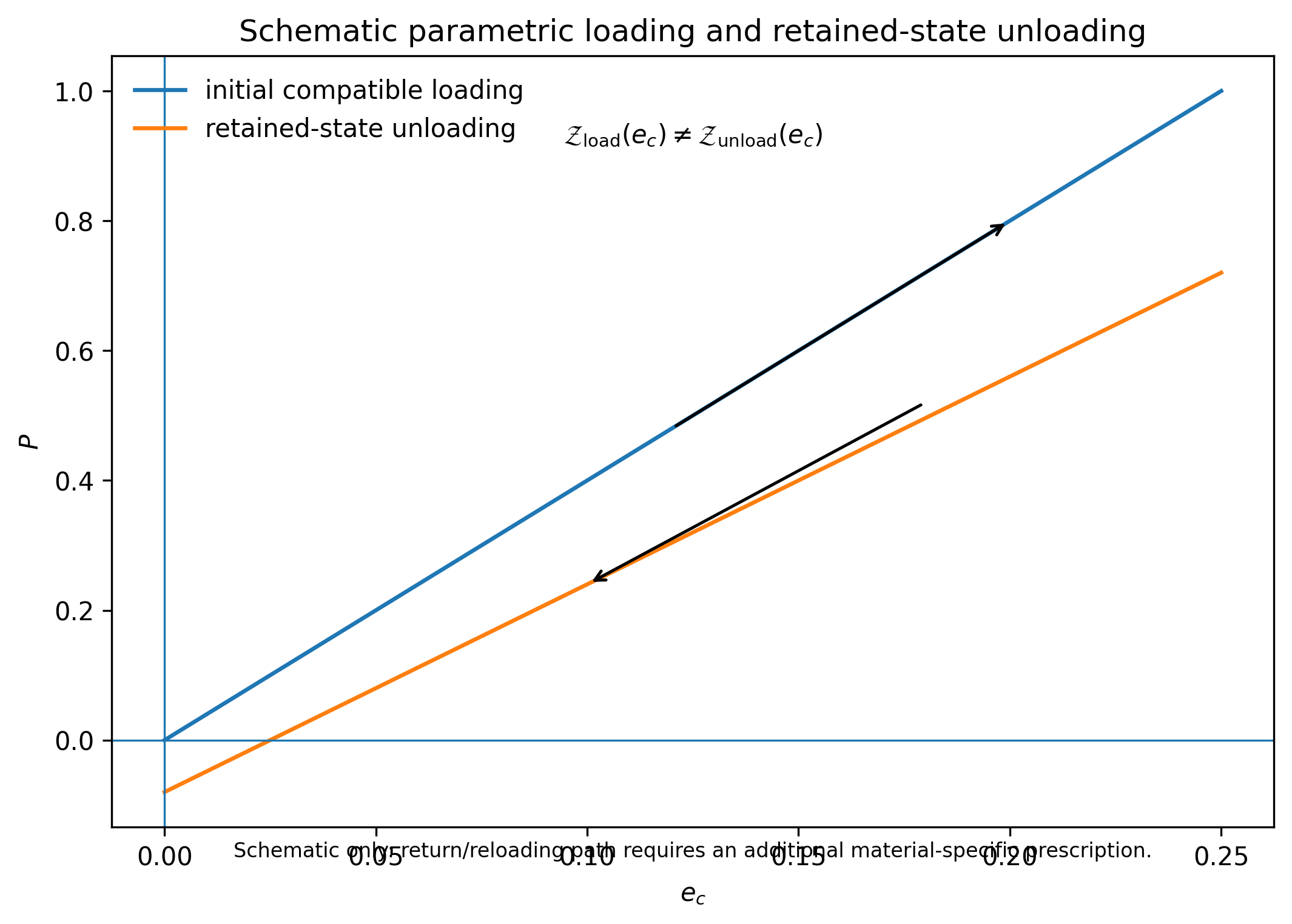}
\caption{Schematic parametric loading and retained-state unloading. Distinct
combined states may occur at the same compatible strain, so the two macroscopic
branches need not coincide. The complete return and reloading path is intentionally
left unspecified because it requires additional material-specific constitutive
information.}
\label{fig:parametric-hysteresis}
\end{figure}

Figure~\ref{fig:parametric-hysteresis} illustrates this distinction schematically.
It is not a material-calibrated closed hysteresis loop: the complete return and
reloading branches require an additional material-specific path prescription.

For the retained state used in Examples 2 and 3,

\begin{equation}
T_{d,r}=1.50807,\qquad S_r=0.35697,
\end{equation}

a branch on which the retained defect structure remains fixed satisfies

\begin{equation}
P_{\mathrm{unload}}=\left(E-\gamma_2T_{d,r}^2\right)e_c-
\alpha T_{d,r}^2,
\end{equation}

or numerically

\begin{equation}
P_{\mathrm{unload}}=77.2572\,e_c-0.22743.
\end{equation}

This differs from the initial compatible branch

\begin{equation}
P_{\mathrm{load}}=Ee_c=100e_c.
\end{equation}

Hence the same value of $e_c$ may correspond to different values of $P$,
depending on the cumulative structural path by which that state has been
reached.

For a closed parametric cycle, the signed area enclosed by the macroscopic
response curve may be written

\begin{equation}
W_{\mathrm{loop}}=\oint P\,de_c.
\end{equation}

A nonzero value $W_{\mathrm{loop}}\neq0$ characterizes macroscopic hysteresis
of the parametric response. Within the present theory this geometrical loop
area is not automatically identified with physical dissipation. Such an
interpretation would require an additional constitutive theory governing the
evolution of the internal structural components. Here the loop expresses the
path dependence of the combined material state and the fact that reversal of
the external parameter does not necessarily reverse the retained structural
state.

\paragraph{Interpretive consequence}

Example 6 therefore describes cumulative hysteresis as a property of a
parametric loading path through the combined state space. It does not
introduce relaxation times or a separate kinetic law. A material-specific
evolution law would be required to determine the detailed traversal of the
path, its rate dependence, or a dissipative interpretation of the loop area.

\subsection{Structural Stabilization and Macroscopic Tangent Softening}

\paragraph{Purpose}

The purpose of this example is to distinguish two related but different
notions: resistance to further structural evolution and the tangent
stiffness of the observable force--strain response. In the cumulative framework
these quantities are generated by the same stored energy, but they need not
change in the same direction. This distinction is especially important after
removing the former threshold parameter $H_s$.

\paragraph{Structural resistance from the cumulative energy}

For the reduced stored energy used in Examples 1--6, the structural tangent
coefficients are

\begin{equation}
K_{TT} = a_T-2\alpha e_c-\gamma_2e_c^2 +3b_TT_d^2-\gamma_3S^2,
\end{equation}

\begin{equation}
K_{TS} = -2\gamma_3T_dS,
\end{equation}

and

\begin{equation}
K_{SS} = \beta_S-\gamma_3T_d^2+3b_SS^2.
\end{equation}

The positive quartic coefficients $b_T$ and $b_S$ increase the diagonal
structural stiffnesses as the corresponding amplitudes grow. In this sense
they provide structural stabilization against unlimited growth. Conversely, the
coupling terms proportional to $\gamma_2$ and $\gamma_3$ can reduce a
structural stiffness and promote further participation of another component.

Thus the same potential contains competing tendencies:

\begin{equation}
\text{structural stabilization through }b_T,b_S, \qquad \text{coupling-induced
reduction through }\gamma_2,\gamma_3.
\end{equation}

\paragraph{Macroscopic tangent on the compatible--defect branch}

On the branch $S=0$, the defect amplitude satisfies

\begin{equation}
T_d^2 = \frac{ 2\alpha e_c+\gamma_2e_c^2-a_T }{b_T}.
\end{equation}

The observable tangent modulus is

\begin{equation}
E_{\mathrm{tan}}^{(12)} = E-\gamma_2T_d^2 - \frac{ 2\left(\alpha+\gamma_2e_c%
\right)^2 }{b_T}.
\end{equation}

For the positive parameters adopted in the numerical example, both
corrections to $E$ are nonpositive. Therefore

\begin{equation}
E_{\mathrm{tan}}^{(12)}<E,
\end{equation}

once the defect component participates. The virgin monotone specialization
used in Examples 1--4 consequently predicts macroscopic softening relative
to the initial compatible stiffness.

This conclusion is not in conflict with the increase of $K_{TT}$ caused by
the $3b_TT_d^2$ term. The former is an observable force--strain tangent
after the internal variable has adjusted along its equilibrium branch; the
latter is the local resistance to a structural perturbation at fixed $e_c$.

\paragraph{Macroscopic tangent on the three-component branch}

Let

\begin{equation}
D_3 = b_T-\frac{\gamma_3^2}{b_S}.
\end{equation}

On the three-component equilibrium branch,

\begin{equation}
E_{\mathrm{tan}}^{(123)} = E-\gamma_2T_d^2 - \frac{ 2\left(\alpha+%
\gamma_2e_c\right)^2 }{D_3}.
\end{equation}

The boundedness condition

\begin{equation}
b_Tb_S>\gamma_3^2
\end{equation}

implies $D_3>0$. Hence, for the present sign convention and positive
parameters,

\begin{equation}
E_{\mathrm{tan}}^{(123)}<E.
\end{equation}

The additional source participation can therefore produce a further
reduction of the macroscopic tangent even while the quartic source term $%
b_SS^4/4$ increases the local resistance to unlimited source growth.

\paragraph{Other macroscopic tangent responses}

The preceding result belongs to the particular reduced potential selected
for the illustrative examples. The cumulative framework itself does not require
macroscopic softening. A hardening-type observable response may arise when
the constitutive energy contains couplings for which the evolving structure
increases the effective compatible stiffness, or when an initially retained
structural component decreases under the applied loading so that the
compatible tangent recovers toward its perfect-state value.

In the full three-dimensional theory, analogous changes may result from the
tensorial couplings among

\begin{equation}
e_c,\quad \omega_c,\quad e_d,\quad \omega_d,\quad T_d,\quad \Omega_d,\quad
T,\quad \Omega,\quad S,\quad R.
\end{equation}

Accordingly, ``hardening'' and ``softening'' are not introduced as separate
macroscopic constitutive branches. They describe the observable consequence
of how the cumulative structural state changes and how that state enters the
total energy.

\paragraph{Tangent-modulus interpretation}

It is useful to distinguish three tangent objects already encountered in
Example 1:

\begin{equation}
\mathbf{K}_{\mathrm{cum}}, \qquad \mathbf{K}_{\mathrm{str}}, \qquad
E_{\mathrm{tan}}=\frac{dP}{de_c}.
\end{equation}

The first tests stability of the complete reduced state under the selected
loading control. The second measures local resistance to structural
perturbations at fixed compatible strain. The third is the directly
observable slope of the macroscopic loading curve after internal
participation has been taken into account.

An increase of a structural tangent coefficient is therefore not
automatically equivalent to an increase of the macroscopic tangent modulus.
The distinction is one of the useful consequences of the cumulative
description.

\paragraph{Numerical specialization}

For the parameter set used in Examples 1--4,

\begin{equation}
E=100,\qquad \alpha=0.1,\qquad \gamma_2=10,\qquad b_T=0.185,
\end{equation}

the compatible branch has

\begin{equation}
E_{\mathrm{tan}}^{(1)}=100.
\end{equation}

Representative values reported previously are

\begin{equation}
E_{\mathrm{tan}}(0.13)=74.7568
\end{equation}

on the compatible--defect branch and

\begin{equation}
E_{\mathrm{tan}}(0.22)=16.8000
\end{equation}

on the three-component branch. For this illustrative specialization the
progressive participation of the internal structure therefore produces a
pronounced softening of the observable response.

These values should not be interpreted as a universal prediction that defect
or source evolution must soften a solid. They demonstrate one constitutive
regime of the cumulative framework. Other choices of the total energy may
produce stiffening, mixed hardening--softening, or nonmonotone tangent
behavior.

\paragraph{Consequences for the illustrative figures}

The former comparison based on the parameter $H_s$ belonged to the previous
threshold evolution law and is not retained. In particular, the old panel
comparing $H_s=-20,1,15$ should not be used as evidence for the present
formulation. Any revised figure should instead display quantities derived
from the current energy, for example

\begin{equation}
P(e_c),\qquad T_d(e_c),\qquad S(e_c),\qquad E_{\mathrm{tan}}(e_c),
\end{equation}

along the equilibrium branches of Examples 1--4.

\paragraph{Structural stabilization and macroscopic tangent response.}

The cumulative formulation distinguishes local structural resistance from
the observable macroscopic tangent. Positive quartic terms may stabilize the
structural directions while coupling to evolving internal variables reduces
the force--strain tangent. For the particular parameter set used below the
observable response softens as structural participation increases. Other
constitutive choices within the cumulative framework may produce hardening,
mixed hardening--softening, or nonmonotone behavior.

\paragraph{Interpretive consequence}

The cumulative energy distinguishes structural stabilization from the
observable macroscopic tangent response. Positive quartic coefficients
increase the local resistance to further growth of defect and source
amplitudes, while the present coupling terms reduce the observable
force--strain tangent as structural participation increases. Thus the
selected numerical specialization predicts macroscopic tangent softening
together with stabilization of the activated structural components. More
generally, hardening or softening is determined by the constitutive
couplings and by the changing participation of the cumulative components,
not by inserting separate phenomenological branches.

\subsection{Spatial Extension, Size Dependence, and Localization Criterion}

\paragraph{Purpose}

The preceding examples are homogeneous and therefore contain no intrinsic
specimen-size dependence. The purpose of the final example is to add the
leading gradient terms consistently to the same cumulative energy and to
examine the stability of a homogeneous combined state with respect to
spatially nonuniform structural perturbations.

The calculation is deliberately restricted to the onset problem. It
identifies the size dependence and neutral condition of admissible nonuniform modes and the
relative defect--source content of such modes. It does not determine a
finite postcritical localization band.

\paragraph{Spatial extension of the same cumulative energy}

Let

\begin{equation}
W_{\mathrm{hom}}(e_c,T_d,S) = \Psi(e_c,T_d,S)
\end{equation}

denote the homogeneous stored energy used in the illustrations. Its simplest
one-dimensional spatial extension is

\begin{equation}
\Pi_L = \int_0^L \left[ W_{\mathrm{hom}}(e_c,T_d,S) + \frac{h_T}{2}%
\left(T_d^{\prime}\right)^2 + \frac{h_S}{2}\left(S^{\prime}\right)^2 \right]%
dx ,
\end{equation}

where

\begin{equation}
h_T>0,\qquad h_S>0.
\end{equation}

The gradient coefficients do not create additional structural components.
They penalize spatial variation of the defect and source amplitudes already
present in the combined state and introduce internal length scales into
the stability problem.

\paragraph{Second variation about a homogeneous state}

Let

\begin{equation}
^* = \left( e_c^*,T_d^*,S^* \right)
\end{equation}

be a homogeneous equilibrium state. Consider structural perturbations

\begin{equation}
T_d \longrightarrow T_d^*+\varepsilon\,\delta T_d, \qquad S \longrightarrow
S^*+\varepsilon\,\delta S .
\end{equation}

At fixed compatible strain, the second variation is

\begin{equation}
\begin{split}
\delta^2\Pi_L = \int_0^L \bigg[ &A_T\left(\delta T_d\right)^2 +2C_{TS}\delta
T_d\delta S +A_S\left(\delta S\right)^2 \\
&+h_T\left(\delta T_d^{\prime}\right)^2 +h_S\left(\delta S^{\prime}\right)^2 %
\bigg]dx ,
\end{split}%
\end{equation}

where

\begin{equation}
A_T = a_T-2\alpha e_c^* -\gamma_2\left(e_c^*\right)^2
+3b_T\left(T_d^*\right)^2 -\gamma_3\left(S^*\right)^2,
\end{equation}

\begin{equation}
A_S = \beta_S-\gamma_3\left(T_d^*\right)^2 +3b_S\left(S^*\right)^2,
\end{equation}

and

\begin{equation}
C_{TS} = -2\gamma_3T_d^*S^*.
\end{equation}

These coefficients are exactly the entries of the structural tangent derived
from the homogeneous cumulative energy. No new local localization energy is
introduced.

\paragraph{Finite-length structural mode}

For an admissible perturbation that vanishes at $x=0,L$, the lowest
nonuniform mode has

\begin{equation}
q_1=\frac{\pi}{L},
\end{equation}

and may be written

\begin{equation}
\delta T_d = A_T^{(1)}\sin\left(\frac{\pi x}{L}\right), \qquad \delta S =
A_S^{(1)}\sin\left(\frac{\pi x}{L}\right).
\end{equation}

The corresponding spatial structural tangent is

\begin{equation}
\mathbf{K}_{\mathrm{loc}}(L) = 
\begin{pmatrix}
A_T+h_T\pi^2/L^2 & C_{TS} \\ 
C_{TS} & A_S+h_S\pi^2/L^2%
\end{pmatrix}%
.
\end{equation}

The neutral condition for this nonuniform mode is

\begin{equation}
\left( A_T+h_T\frac{\pi^2}{L^2} \right) \left( A_S+h_S\frac{\pi^2}{L^2}
\right) - C_{TS}^2 = 0 .
\end{equation}

The critical eigenvector gives the relative defect and source participation
through

\begin{equation}
\frac{A_S^{(1)}}{A_T^{(1)}} = - \frac{ A_T+h_T\pi^2/L^2 }{ C_{TS} },
\end{equation}

when $C_{TS}\neq0$. An equivalent expression follows from the second row of
the neutral system.

\paragraph{Perfect-branch spatial defect mode}

On the perfect branch,

\begin{equation}
T_d^*=0,\qquad S^*=0,
\end{equation}

so that

\begin{equation}
C_{TS}=0, \qquad A_S=\beta_S,
\end{equation}

and

\begin{equation}
A_T = a_T-2\alpha e_c-\gamma_2e_c^2.
\end{equation}

The first admissible nonuniform defect mode satisfies

\begin{equation}
a_T-2\alpha e_c-\gamma_2e_c^2 + h_T\frac{\pi^2}{L^2} = 0 .
\end{equation}

Hence its critical compatible strain is

\begin{equation}
e_{c,\mathrm{sp}}(L) = \frac{ -\alpha+ \left[ \alpha^2+ \gamma_2 \left(
a_T+h_T\pi^2/L^2 \right) \right]^{1/2} }{ \gamma_2 }.
\end{equation}

As $L$ decreases, the positive gradient contribution grows and the
nonuniform defect mode requires a larger compatible strain. Thus the spatial
extension predicts a direct size dependence for this constrained nonuniform
perturbation.

In the limit

\begin{equation}
L\longrightarrow\infty,
\end{equation}

the gradient contribution vanishes and

\begin{equation}
e_{c,\mathrm{sp}}(L) \longrightarrow e_{c,2},
\end{equation}

where $e_{c,2}$ is the homogeneous defect bifurcation strain of Example 1.

\paragraph{Spatial source mode on the compatible--defect branch}

On the homogeneous compatible--defect branch,

\begin{equation}
S^*=0,\qquad T_d^*\neq0,
\end{equation}

and therefore

\begin{equation}
C_{TS}=0.
\end{equation}

The source-direction coefficient is

\begin{equation}
A_S = \beta_S-\gamma_3\left(T_d^*\right)^2.
\end{equation}

The first admissible nonuniform source mode satisfies

\begin{equation}
\beta_S -\gamma_3\left(T_d^*\right)^2 + h_S\frac{\pi^2}{L^2} = 0,
\end{equation}

or

\begin{equation}
\left(T_d^*\right)^2 = \frac{ \beta_S+h_S\pi^2/L^2 }{ \gamma_3 }.
\end{equation}

The gradient term therefore delays the nonuniform source mode relative to
the homogeneous source bifurcation

\begin{equation}
T_{d,3}^2=\frac{\beta_S}{\gamma_3}.
\end{equation}

Again the finite-length contribution approaches the homogeneous result as $L$
becomes large.

\paragraph{Three-component branch}

When

\begin{equation}
T_d^*\neq0,\qquad S^*\neq0,
\end{equation}

the equilibrium equations simplify the structural tangent coefficients to

\begin{equation}
A_T = 2b_T\left(T_d^*\right)^2,
\end{equation}

\begin{equation}
A_S = 2b_S\left(S^*\right)^2,
\end{equation}

and

\begin{equation}
C_{TS} = -2\gamma_3T_d^*S^*.
\end{equation}

The homogeneous structural determinant is therefore

\begin{equation}
A_TA_S-C_{TS}^2 = 4\left(T_d^*\right)^2\left(S^*\right)^2 \left(
b_Tb_S-\gamma_3^2 \right).
\end{equation}

Because the boundedness condition requires

\begin{equation}
b_Tb_S>\gamma_3^2,
\end{equation}

this determinant is positive on a nonzero stable three-component branch.
Adding positive gradient terms gives

\begin{equation}
\begin{split}
\det\mathbf{K}_{\mathrm{loc}}(L) ={}& A_TA_S-C_{TS}^2 +A_Th_S\frac{\pi^2}{L^2} +A_Sh_T%
\frac{\pi^2}{L^2} \\
&+ h_Th_S\frac{\pi^4}{L^4},
\end{split}%
\end{equation}

which is still positive.

This result has an important interpretation. The present minimal
positive-gradient extension regularizes spatial perturbations but does not,
by itself, produce a finite-wavelength localization instability of an
already stable three-component equilibrium branch. The gradient terms
stabilize spatial variation; they do not create it.

\paragraph{Meaning of a localization prediction}

A genuine finite-width localization calculation requires, in addition to
gradient regularization, a constitutive mechanism that makes the local
homogeneous response unstable or nonconvex in the relevant structural
direction. The gradient terms can then select a length scale and regularize
the emerging spatial profile.

Such a mechanism could arise from a more detailed constitutive dependence of
the total cumulative energy, from state-dependent gradient coefficients, or
from a post-bifurcation evolution law. Introducing one of those mechanisms
would be a further constitutive specialization and is not required for the
general cumulative framework developed in this paper.

Accordingly, Example 8 establishes the spatial stability operator and its
size dependence but does not claim a finite-wavelength localization band
from a model whose local structural energy remains stable.

\paragraph{Relation to classical stability}

This construction is mathematically analogous to the energy and eigenvalue
criterion used in classical elastic stability \cite{TimoshenkoGere1961}. In
both cases a critical state is identified by loss of positive definiteness
of a second variation. The difference is the nature of the neutral
perturbation. In classical buckling it is typically a displacement mode,
whereas here it is a perturbation in the defect and source components of the
cumulative structural state.

The analogy concerns the variational criterion, not an identification of
structural localization with geometric buckling.

\paragraph{Relation to Cosserat localization models}

Recent deformable-Cosserat formulations provide other mechanisms by which
additional continuum structure can regularize localization. Panteghini and
Rubin separate dissipative macro-continuum response from an energetic
micro-continuum contribution and obtain an intrinsic length scale governing
localization patterns \cite{PanteghiniRubin2026}. The present calculation
has a different purpose. It shows how gradient terms associated with the
defect and source components enter the second variation of the cumulative
energy and how the finite-size contribution modifies the structural
stability operator.

\paragraph{Size dependence and spatial stability.}

Gradient terms in $T_d$ and $S$ introduce internal length scales into the
second variation. For constrained nonuniform modes, smaller specimens carry
larger gradient penalties and therefore require larger deformation to reach
the corresponding neutral condition. Example 8 derives this size dependence
directly from the spatial extension of the same stored energy.

\paragraph{Localization.}

The present minimal positive-gradient model provides a spatial stability
operator and regularizes nonuniform structural perturbations. On an already
stable three-component branch, however, the positive gradient terms do not
by themselves generate a finite-wavelength localization instability. A true
post-bifurcation localization band requires an additional local softening
mechanism, state-dependent spatial constitutive law, or postcritical
evolution equation. Localization is therefore discussed only to the extent
supported by the selected energy.

These interpretations are deliberately narrower than claims that the
cumulative framework automatically reproduces every familiar inelastic
phenomenon. The illustrative examples instead show how several such
responses can be organized within one structural state and also identify
which predictions require further constitutive specialization.

\paragraph{Interpretive consequence}

The spatial extension introduces internal length scales through $h_T$ and $%
h_S$. For admissible nonuniform modes, smaller specimens carry larger
gradient penalties and therefore require a larger deformation to reach the
corresponding neutral condition. The critical eigenvector determines the
relative defect and source content of a coupled spatial mode.

For the particular stable quartic energy used in the illustrations, the
positive gradient terms do not independently generate localization on the
stable three-component branch. They regularize spatial perturbations. A true
finite-wavelength localization prediction requires an additional local
softening or post-bifurcation constitutive mechanism. This conclusion
separates what follows rigorously from the present cumulative energy from
what would require a further material specialization.

\paragraph{Summary of the illustrative examples}

The eight illustrative examples demonstrate distinct capabilities of the
same cumulative framework. Examples 1--4 show structural bifurcation,
retention, residual deformation, residual stress under constraint, and
dependence on the initial structural state. Example 5 shows how retained
structure shifts reverse loading. Example 6 shows parametric cumulative hysteresis produced by path-dependent
retention of the combined structural state. Example 7 distinguishes
structural stabilization from the observable macroscopic tangent response. Example 8
introduces the spatial stability operator and the associated size dependence.

The examples are not calibrated to a particular material, and no claim is
made that all material behavior is exhausted by the selected reduced energy.
Their purpose is to show how apparently different macroscopic responses can
be organized as consequences of the changing participation and coupling of
the cumulative structural components.

\subsection{Numerical and Graphical Illustrations}

The following nondimensional parameter set is used only to illustrate the
structure of the reduced model:

\begin{equation}
E=100,\qquad \alpha=0.1,\qquad \gamma_2=10,\qquad a_T=0.12,
\end{equation}

\begin{equation}
b_T=0.185,\qquad \beta_S=0.1,\qquad \gamma_3=0.1,\qquad b_S=1.
\end{equation}

The quartic stability requirement is satisfied because

\begin{equation}
b_Tb_S=0.185>\gamma_3^2=0.01.
\end{equation}

For the virgin initial state,

\begin{equation}
e_{c,2}=0.10000,\qquad P_2=10.0000,
\end{equation}

and

\begin{equation}
e_{c,3}=0.16493,\qquad P_3=14.7436.
\end{equation}

\begin{figure}[htbp]
\centering \includegraphics[width=0.82%
\textwidth]{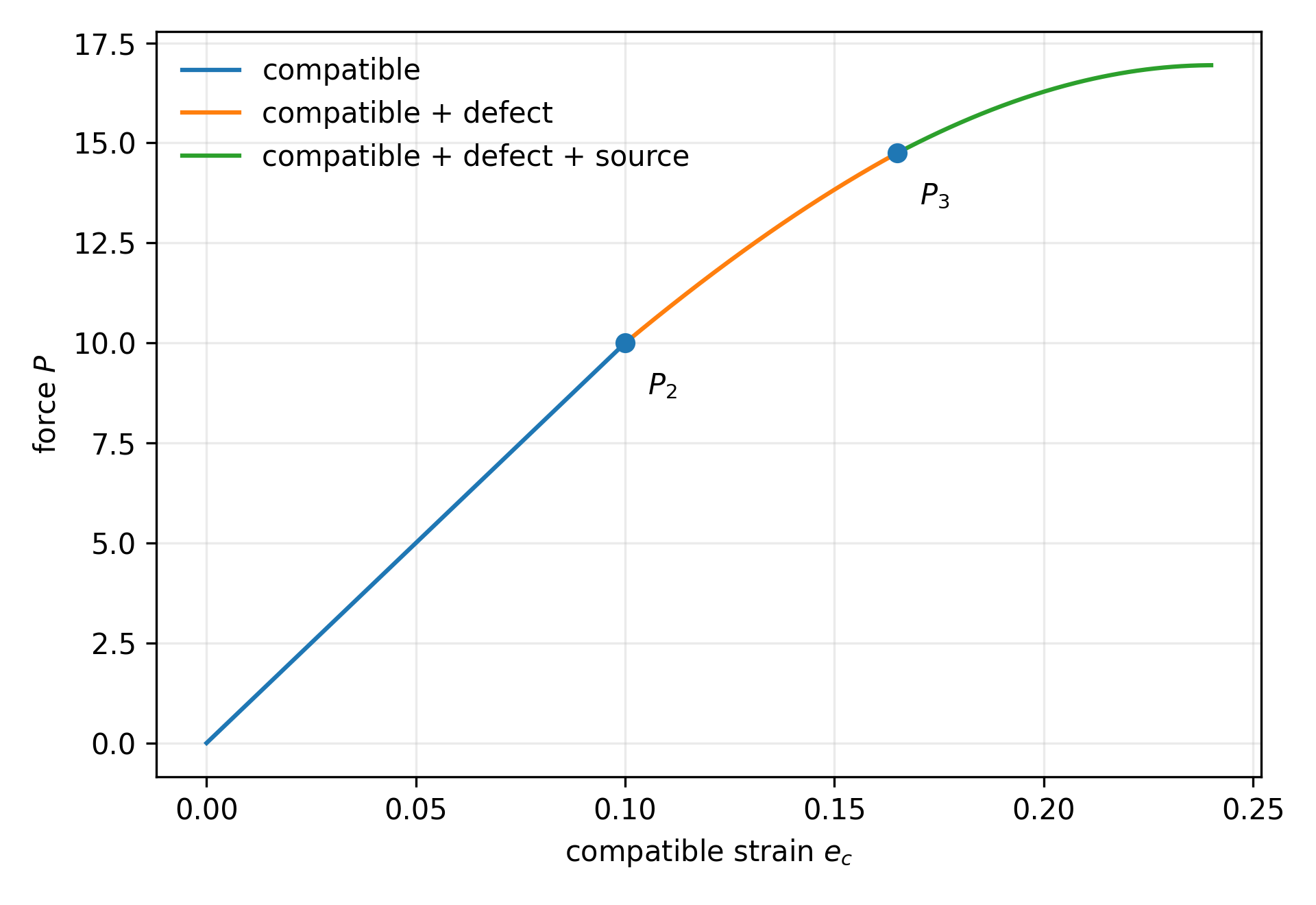}
\caption{One possible path-dependent structural activation for the illustrative
nondimensional parameter set and a perfect initial state. Compatible response is
followed by participation of distributed incompatibility and then defect-source
structure. The points $P_2$ and $P_3$ mark the corresponding bifurcations; the
sequence is not a universal hierarchy of the theory.}
\label{fig:cumulative-loading}
\end{figure}

Figure~\ref{fig:cumulative-loading} shows the implementation of the material
response strategy during loading. The successive branches represent
cumulative participation, $Z_1\longrightarrow Z_1+Z_2\longrightarrow
Z_1+Z_2+Z_3$, rather than replacement of one structural level by another.

For the retained state selected at $e_{c,\max}=0.22$,

\begin{equation}
T_{d,r}=1.50807,\qquad S_r=0.35697,
\end{equation}

and

\begin{equation}
e_{c,r}=0.002944,\qquad P_r=-\alpha T_{d,r}^2=-0.22743.
\end{equation}

\begin{figure}[htbp]
\centering \includegraphics[width=0.82%
\textwidth]{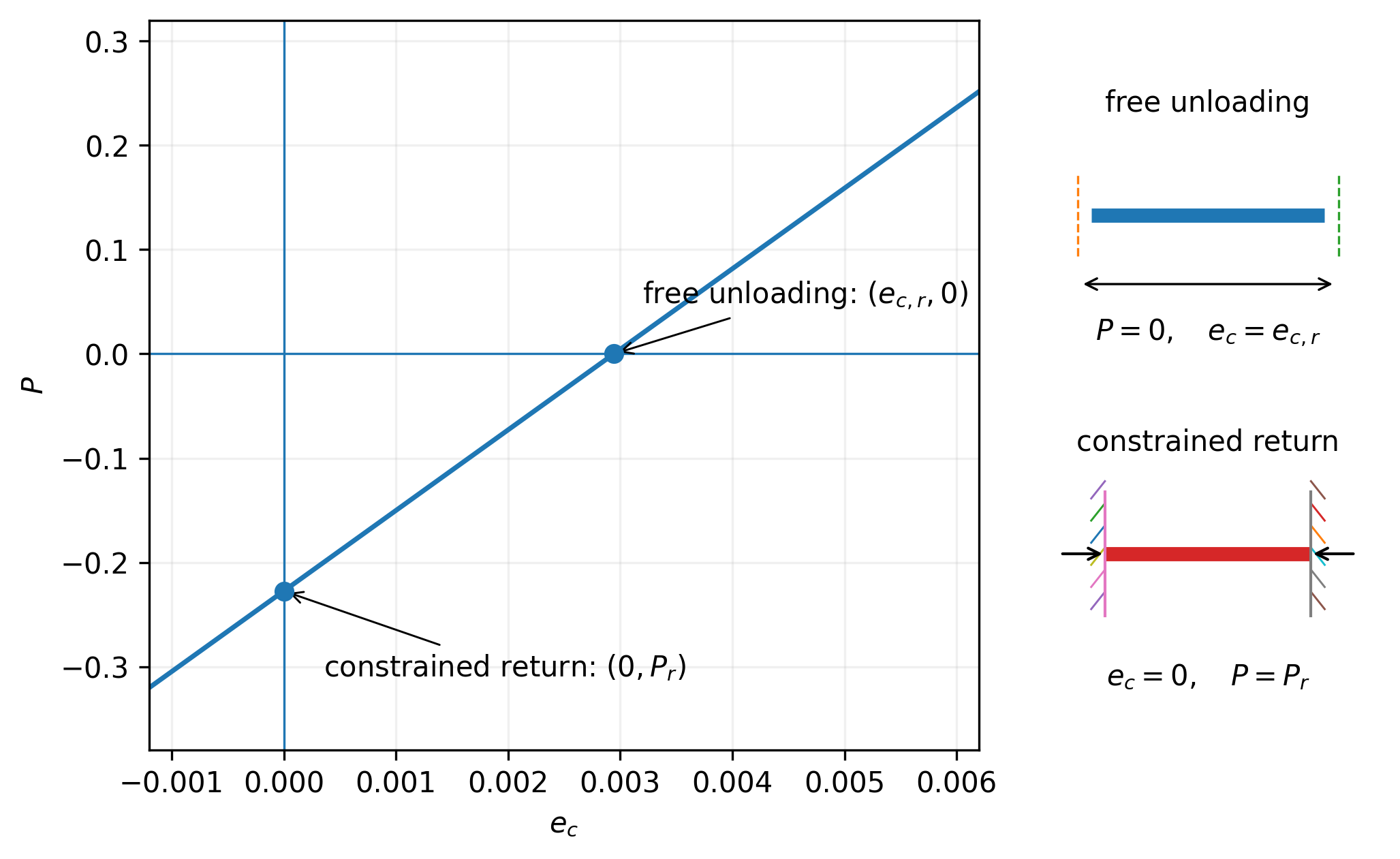}
\caption{Residual deformation and residual stress generated by the same
retained combined structural state. When the external boundary conditions
return to their initial zero-load values, $P(\protect\lambda)=0$ at $%
e_c=e_{c,r}$. If compatible return is instead constrained by $e_c=0$, the
retained state produces the nonzero residual reaction $P(\protect\lambda)=P_r
$.}
\label{fig:residual-strain-stress}
\end{figure}

Figure~\ref{fig:residual-strain-stress} displays the two boundary
manifestations of the same retained state: residual deformation when the
external boundary conditions return to their initial zero-load values and
residual stress when compatible return is constrained.

\begin{figure}[htbp]
\centering \includegraphics[width=0.82%
\textwidth]{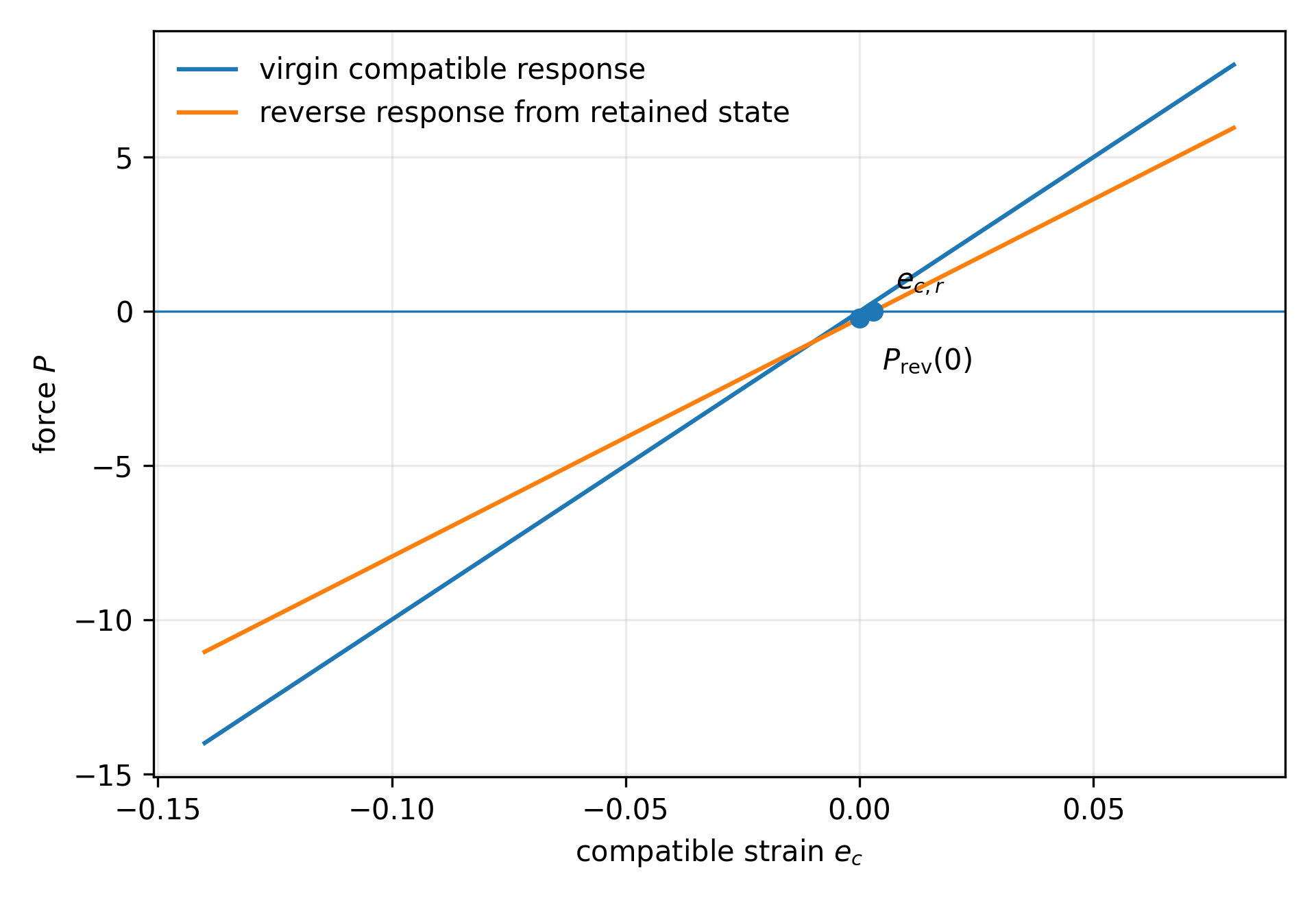}
\caption{Bauschinger-type directional memory produced by a retained combined
state. The retained defect amplitude changes both the intercept and tangent of the
reverse response; no unique reverse reactivation threshold is implied without an
additional evolution law.}
\label{fig:reverse-response}
\end{figure}

Figure~\ref{fig:reverse-response} compares the virgin compatible response
with reverse loading from the same representative retained state. The resulting
shift is interpreted as Bauschinger-type directional memory; no unique reactivation
load is implied without a specified evolution law.

Example 6 is formulated parametrically and does not introduce a separate
kinetic specialization. The retained-state unloading branch used there is

\begin{equation}
P_{\mathrm{unload}}=77.2572\,e_c-0.22743.
\end{equation}

A closed numerical hysteresis loop is not assigned here because its detailed
return and reloading branches require an additional material-specific path or
evolution prescription. Figure~\ref{fig:parametric-hysteresis} therefore remains
a schematic representation of the path-dependent distinction between loading and
retained-state unloading.

The observable tangent modulus on the virgin equilibrium branches takes the
representative values

\begin{equation}
E_{\mathrm{tan}}(0.05)=100.0000,\qquad E_{\mathrm{tan}}(0.13)=74.7568,\qquad
E_{\mathrm{tan}}(0.22)=16.8000.
\end{equation}

\begin{figure}[htbp]
\centering \includegraphics[width=0.82%
\textwidth]{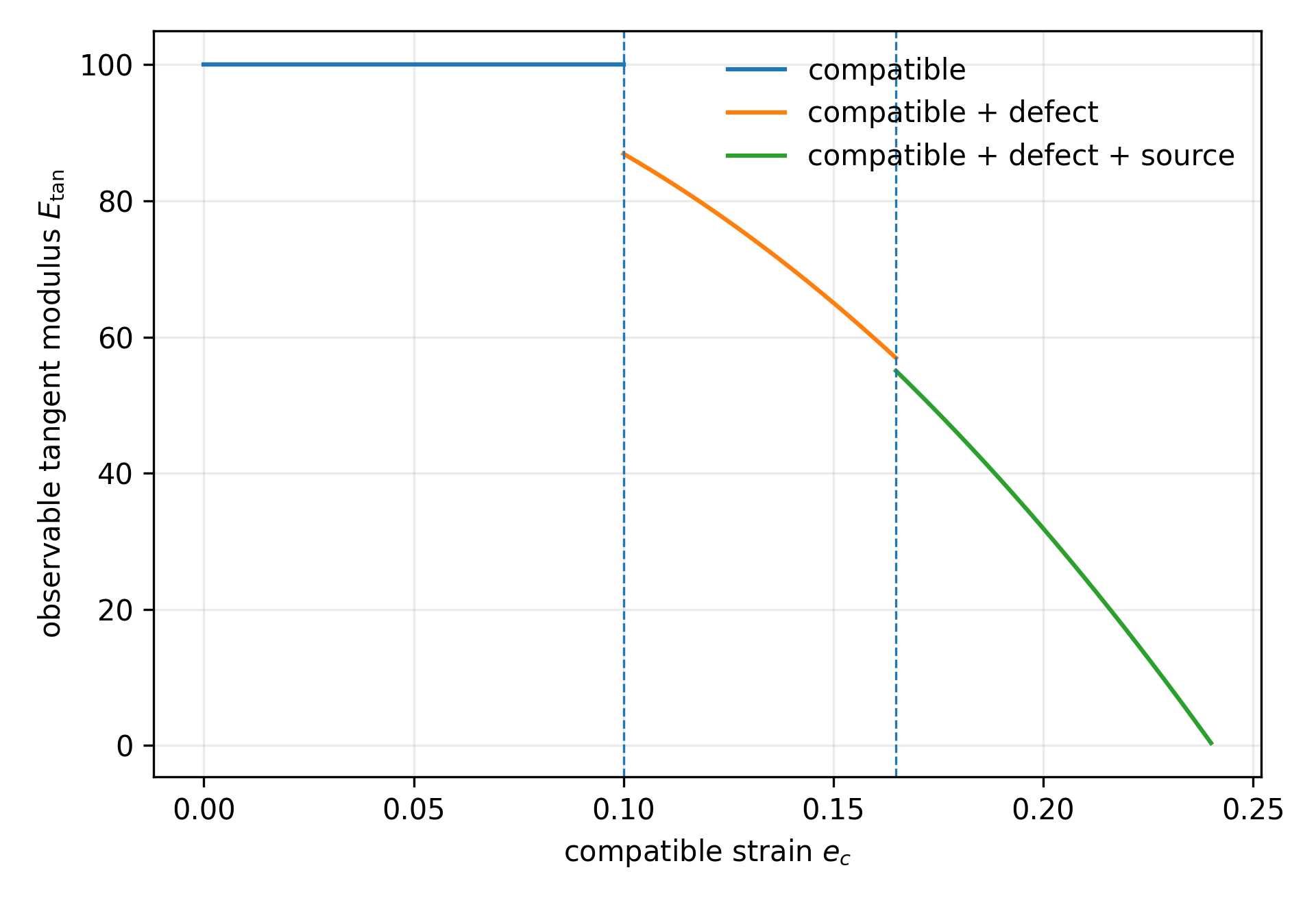}
\caption{Observable tangent modulus along the virgin loading path. The selected
parameter set illustrates macroscopic tangent softening as the activated internal
structural components remain stabilized by the higher-order terms of the stored
energy.}
\label{fig:tangent-modulus}
\end{figure}

Figure~\ref{fig:tangent-modulus} shows the branchwise change in tangent
stiffness. The softening belongs to this constitutive specialization and is
not a universal requirement of the cumulative framework.

Finally, define the normalized inverse-size parameter

\begin{equation}
\lambda_T=\frac{h_T\pi^2}{a_TL^2}.
\end{equation}

Then

\begin{equation}
\frac{e_{c,\mathrm{sp}}}{e_{c,2}} = \frac{-\alpha+\sqrt{\alpha^2+%
\gamma_2a_T(1+\lambda_T)}} {-\alpha+\sqrt{\alpha^2+\gamma_2a_T}}.
\end{equation}

\begin{figure}[htbp]
\centering \includegraphics[width=0.82%
\textwidth]{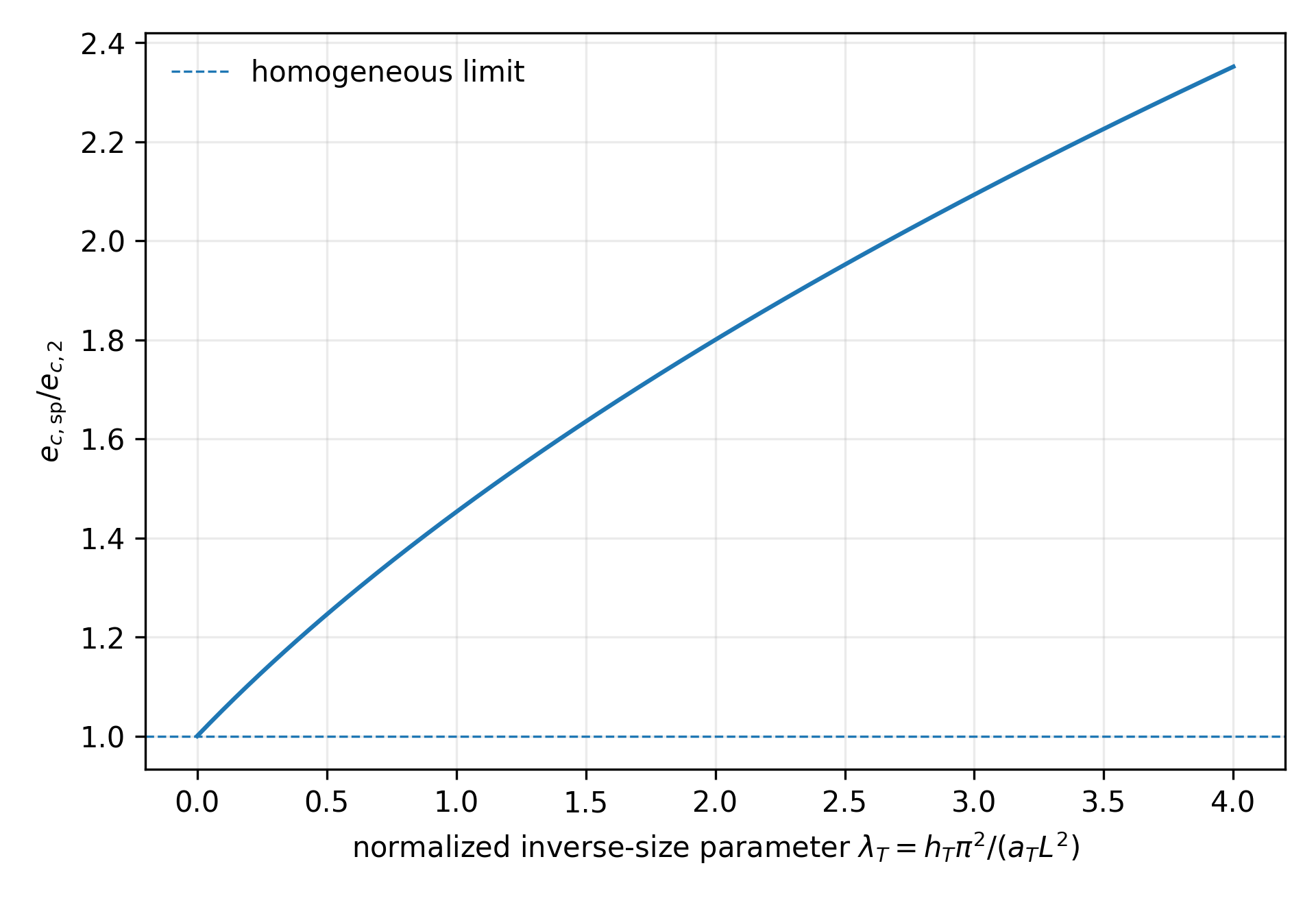}
\caption{Normalized parametric size effect for the first nonuniform defect
mode. The homogeneous limit is recovered at $\protect\lambda_T=0$.}
\label{fig:size-effect}
\end{figure}

Figure~\ref{fig:size-effect} presents the resulting normalized parametric
size effect. Because $h_T$ and $h_S$ are not calibrated, the curve
illustrates the predicted qualitative size dependence rather than a
material-specific quantitative fit.

Taken together, the illustrations demonstrate structural bifurcation,
residual deformation and stress, initial-state dependence, reverse-response
asymmetry, parametric cumulative hysteresis, changes of tangent stiffness,
and spatial size dependence. The examples remain
illustrative rather than material-calibrated.

\clearpage

\section{Discussion and Conclusions}

The present work develops a cumulative framework for solid deformation in which
compatible deformation, distributed incompatibility, and defect-source
structure are represented as recursively related components of a single
material structure. At a prescribed value of the externally controlled
parameter $\lambda$, these components form the combined state
\begin{equation}
\mathcal Z(\lambda)=\bigl(\mathcal Z_c(\lambda),\mathcal Z_d(\lambda),
\mathcal Z_s(\lambda)\bigr).
\end{equation}
The components are geometrically distinct but are not independent levels of
material response or separate boundary-value problems. They are coupled
through the common material geometry, cumulative energy, and variational
field equations, and are subject to the same physical boundary and initial
conditions of the body.

The term \emph{cumulative} characterizes the development of this combined
state as the externally affecting conditions vary along a path parametrized
by $\lambda$. The parameter need not represent physical time. Depending on
the initial state, constitutive properties, and loading path, the relative
participation of the three structural components may change through
path-dependent activation and retention. Consequently, the sequence
\begin{equation}
(\mathcal Z_c,0,0)\longrightarrow(\mathcal Z_c,\mathcal Z_d,0)
\longrightarrow(\mathcal Z_c,\mathcal Z_d,\mathcal Z_s)
\end{equation}
is a possible response of an initially perfect body, rather than a universal
hierarchy imposed by the theory. A nonvirgin body may contain distributed
incompatibility or defect-source structure already in its initial combined
state.

The variational formulation provides a common mathematical description of
these recursively coupled components. The compatible component supplies the
underlying Cosserat material geometry; distributed torsion and curvature
characterize incompatibility relative to this retained structure; and
defect-source fields extend the description relative to the coupled
compatible--defect background. The resulting recursion means that
participation of an additional structural component enlarges the combined
state rather than replacing the previously established structure.

The reduced examples illustrate several consequences of this construction.
Successive bifurcation from a perfect initial state demonstrates one possible
path of structural activation. Retention of activated structure produces
residual deformation or residual stress according to the imposed boundary
conditions. Nonvirgin initial states show that the same theory does not
require the virgin activation sequence. Reverse loading exhibits
Bauschinger-type directional memory through a modified mechanical path,
while a quantitative reverse reactivation threshold would require an
additional evolution law. Parametric cyclic loading similarly shows how
different combined states may be reached at the same value of the macroscopic
deformation on loading and unloading, providing a cumulative description of
hysteresis without identifying the loading-path parameter with physical time.

The examples also distinguish internal structural stabilization from the
observable macroscopic response. Higher-order structural terms stabilize the
activated defect and source components, while their participation may
simultaneously reduce the macroscopic tangent modulus. The spatial extension
further shows that positive defect and source gradient terms introduce a
characteristic size dependence and modify the neutral condition for
nonuniform structural modes. For the stable quartic specialization considered
here, these gradient terms regularize spatial variations and do not by
themselves generate a finite-width localization instability; such
localization requires additional constitutive structure capable of producing
the necessary loss of local stability.

The principal contribution is therefore not a prescribed sequence of deformation
mechanisms, but a framework in which different structural characteristics of
a solid are retained within one recursively coupled material state and may
participate differently as external conditions change. The cumulative framework
is constituted by the recursive combination of these structural components,
their common physical boundary and initial conditions, and their
path-dependent participation and retention within the combined material
state.

\end{document}